\documentclass{kaisPichic}
\date{}
\usepackage{srcltx}
\usepackage{multicol}
\usepackage{amsmath}
\usepackage{amssymb}
\usepackage{latexsym}
\usepackage{graphicx}
\usepackage{epsfig}
\usepackage{color}
\usepackage{epstopdf}

\usepackage{enumitem}
\setlist{topsep=3pt}

\usepackage{color}
\usepackage{amsmath,amssymb}
\usepackage{amsfonts}
\usepackage{graphicx}
\usepackage{epsfig}
\usepackage{wrapfig}

\newcommand{\ignore}[1]{}

\newcommand{\re}[1]{\textcolor{red}{#1}}
\newcommand{\bl}[1]{\textcolor{blue}{#1}}

\newcommand{\boxtheorem}{\hfill $\blacksquare$\\}

\newcommand{\nit}[1]{{\it #1}}
\newcommand{\ca}[1]{$\mathcal{ #1}$}
\newcommand{\mc}[1]{\mathcal{ #1}}
\newcommand{\maf}[1]{\mathfrak{ #1}}

\newcommand{\comlb}[1]{{\vspace{2mm}\noindent \bf \re{COMM(LEO):}}~ #1 \hfill {\bf
    END.}\\}

\newcommand{\comfr}[1]{{\vspace{2mm}\noindent \bf \bl{COMM(Flavio):}}~ #1 \hfill {\bf
    END.}\\}

\newtheorem{definition}{Definition}
\newtheorem{example}{Example}

\newcounter{remark-counter}
\newcounter{theorem-counter}
\newcounter{corollary-counter}
\newcounter{lemma-counter}
\newcounter{definition-counter}
\newcounter{example-counter}
\newcounter{proposition-counter}
\newcounter{lemmaA-counter}
\newcounter{propA-counter}
\newcounter{defA-counter}

{ \refstepcounter{lemmaA-counter}%
\noindent {\bf Lemma A.\arabic{lemmaA-counter}.}}%

{\refstepcounter{propA-counter}%
\noindent {\bf Proposition A.\arabic{propA-counter}.}}%

{\refstepcounter{defA-counter}%
\noindent {\bf Definition A.\arabic{defA-counter}.}}%

\newcommand {\out}[1]{}

\begin{document}

\title{\bf Contexts and Data Quality Assessment}

\author{{\bf Leopoldo Bertossi}\thanks{Corresponding author. Email: bertossi@scs.carleton.ca} \ \ and \ \ {\bf Flavio Rizzolo}\\Carleton University\\
School of Computer Science\\ Ottawa, Canada.}

\maketitle

\begin{abstract}
The quality of data is context dependent. Starting from this
intuition and experience, we propose and develop a conceptual
framework that captures in formal terms the notion of {\em context-dependent data quality}. We start by proposing a generic and abstract notion of
context,  and also of its uses, in general and in data management in
particular. On this basis, we investigate {\em data quality assessment} and
{\em quality query answering} as context-dependent activities. A context
for the assessment of a database $D$ at hand is modeled as an
external database schema, with possibly materialized or virtual
data, and connections to external data sources. The database $D$
is put in context via mappings to the contextual schema, which
produces a collection $\mc{C}$ of alternative clean versions of
$D$. The quality of $D$ is measured in terms of its distance to
$\mc{C}$. The class $\mathcal{C}$ is also used to define and do quality query
answering. The proposed model allows for natural extensions, like
the use of data quality predicates, the optimization of the access
by the context to external data sources, and also the
representation of contexts by means of more expressive ontologies.
\end{abstract}


\vspace{-1mm}
\section{Introduction}\label{sec:intro} \vspace{-3mm}

The assessment of the quality of a data source is context
dependent, i.e. the notions of ``good'' or ``poor'' data cannot be
separated from the context in which the data is produced and used.
For instance, the data about yearly sales of a product with
seasonal variations might be considered quality data by a business
analyst assessing the yearly revenue of a product. However, the
same data may not be good enough for a warehouse manager who is
trying to estimate the orders for  next month.

In addition, data quality is related to the discrepancy between
the actual stored values and the ``real'' values that were
supposed to be stored. For instance, if a temperature measurement
is taken with a faulty thermometer, the stored value (the
measurement) would differ from the right value (the actual
temperature), which was the one supposed to be stored. This is an
example of {\em semantically inaccurate data}
\cite{batini06.data-quality}.

Another type of semantic discrepancy occurs when {\em senses or
meanings} attributed by the different agents to the actual values
in the database disagree \cite{DBLP:conf/er/JiangBM08}, as shown
in the Example \ref{exa:intro}. In this paper, we focus on data
quality (DQ) problems caused by this type of semantic discrepancy.

\begin{example} \label{exa:intro}
Tom is a patient in a hospital. Several times a day his
temperature is measured and recorded by a nurse. His doctor, John,
wants to see Tom's temperature around noon every day to follow his
evolution. The information that John needs appears in the
\textbf{TempNoon} relation of Table~\ref{tab:temp}, which contains
the temperatures between 11:30 and 12:30 per day for each of
John's patients.

\begin{table}
\vspace*{-4mm}
\begin{center}
\hspace*{-5mm}\begin{minipage}{40mm}
{\small \setlength{\tabcolsep}{2pt}
\begin{center}
\begin{tabular}{r|l|r|r|l|}
\multicolumn{1}{l}{} & \multicolumn{4}{c}{\textbf{TempNoon}} \\
\cline{2-5} & \textbf{Patient} & \textbf{Value} & \textbf{Time} &
\textbf{Date}
\\\cline{2-5}
{\tiny 1}& Tom Waits & 38.5 & 11:45 & Sep/5 \\
{\tiny 2}& Tom Waits & 38.2 & 12:10 & Sep/5 \\
{\tiny 3}& Tom Waits & 38.1 & 11:50 & Sep/6 \\
{\tiny 4}& Tom Waits & 38.0 & 12:15 & Sep/6 \\
{\tiny 5}& Tom Waits & 37.9 & 12:15 & Sep/7 \\
\cline{2-5}
\end{tabular}
\caption{} \label{tab:temp}
\end{center}
}
\end{minipage}
\vspace*{-8mm}\hspace{1.7cm}\begin{minipage}{40mm}
{\small \setlength{\tabcolsep}{2pt}
\begin{center}
\begin{tabular}{r|l|r|r|l|}
\multicolumn{1}{l}{} & \multicolumn{4}{c}{\textbf{TempNoon\!{\large '}}} \\
\cline{2-5} & \textbf{Patient} & \textbf{Value} & \textbf{Time} &
\textbf{Date}
\\\cline{2-5}
{\tiny 1}& Tom Waits & 38.5 & 11:45 & Sep/5 \\
{\tiny 2}& Tom Waits & 38.0 & 12:15 & Sep/6 \\
{\tiny 3}& Tom Waits & 37.9 & 12:15 & Sep/7 \\ \cline{2-5}
\end{tabular}
 \vspace{2mm}\caption{} \label{tab:temp1}
\end{center}
}
\end{minipage}
\end{center}
\end{table}

John has additional \emph{quality} requirements for the
temperature measurements of his patients: they have to be taken by
a certified nurse with an oral thermometer. On Sep/5, unaware of
the new requirements, Cathy takes Tom's temperature at 12:10 with
a \emph{tympanal} thermometer and records the result as the tuple
number 2 in Table \ref{tab:temp}. Since the instrument used does
not appear in the table, John interprets the 38.2$^o$C value as
taken with an oral thermometer.

This is an example of a discrepancy between the semantics of the
value as intended by the data producer (38.2$^o$C taken with a
tympanal thermometer) and the semantics expected by the data
consumer (38.2$^o$C taken with an oral thermometer). This
tuple should not appear in a quality table, i.e. one that satisfies
John's quality requirements, since such a table would contain only
temperatures taken with an oral thermometer.

A similar problem appears in the third tuple in
Table~\ref{tab:temp}: it was taken by a new nurse, Helen, who is
not yet certified, and thus, does not satisfy one of the doctor's
requirements. This tuple should not appear in a quality table
containing only temperatures taken by certified nurses.

Table~\ref{tab:temp1} fixes the problems of Table~\ref{tab:temp}
with respect to the doctor's specification: the problematic second
and third tuples do not appear in it.

How can we say or believe that Table \ref{tab:temp1} does contain
only quality data? Prima facie it does not look much different
from Table~\ref{tab:temp}. This positive assessment would be
possible if we had a {\em contextual database} containing
additional information, e.g. Tables~\ref{tab:nurses},
\ref{tab:cert} and~\ref{tab:thermometers}.

\begin{table}
\begin{center}
\hspace*{5mm}\begin{minipage}{40mm}
{\small \setlength{\tabcolsep}{2pt}
\begin{center}
\begin{tabular}{r|l|l|l|}
\multicolumn{1}{l}{} & \multicolumn{3}{c}{\textbf{S} (shift)}\\\cline{2-4}
& \textbf{Date} & \textbf{Shift} & \textbf{Nurse}\\\cline{2-4}
{\tiny 1} & Sep/5 & morning & Susan \\
{\tiny 2} & Sep/5 & afternoon & Cathy \\
{\tiny 3} & Sep/5 & night & Joan  \\
{\tiny 4} & Sep/6 & morning & Helen  \\
{\tiny 5} & Sep/6 & afternoon & Cathy  \\
{\tiny 6} & Sep/6 & night & Cathy \\
{\tiny 7} & Sep/7 & morning & Susan \\
{\tiny 8} & Sep/7 & afternoon & Susan \\
{\tiny 9} & Sep/7 & night & Joan  \\
\cline{2-4}
\end{tabular}
  \vspace{2mm}\caption{} \label{tab:nurses}
\end{center}
}
\end{minipage}
\hspace{36mm}\begin{minipage}{33mm}
{\small \setlength{\tabcolsep}{2pt}
\begin{center}
\begin{tabular}{r|l|c|}
\multicolumn{1}{l}{} & \multicolumn{2}{c}{\textbf{C}
(certification)}\\\cline{2-3} & \textbf{Name} & \textbf{Year}
\\\cline{2-3}
{\tiny 1} & Ann   & 2003 \\
{\tiny 2} & Cathy & 2009 \\
{\tiny 3} & Irene & 2000 \\
{\tiny 4} & Karen & 1995 \\
{\tiny 5} & Nancy & 1995 \\
{\tiny 6} & Natasha & 2001 \\
{\tiny 7} & Susan & 1996 \\
\cline{2-3}
\end{tabular}
  \vspace{2mm}\caption{} \label{tab:cert}
\end{center}
}
\end{minipage}
\end{center}
\end{table}

\begin{table}
\vspace*{-4mm}
\begin{center}
\hspace*{-5mm}
\begin{minipage}{47mm}
{\small \setlength{\tabcolsep}{2pt}
\begin{center}
\begin{tabular}{r|l|l|l|l|}
\multicolumn{1}{l}{} & \multicolumn{4}{c}{\textbf{I}
(instrument)}\\\cline{2-5} & \textbf{Nurse} & \textbf{Date} &
\textbf{Instr} & \textbf{Type} \\\cline{2-5}
{\tiny 1} & Susan & Sep/5 & Therm. & Oral \\
{\tiny 2} & Susan & Sep/5 & BPM    & Arm \\
{\tiny 3} & Cathy & Sep/5 & Therm. & Tymp \\
{\tiny 4} & Cathy & Sep/5 & BPM    & Arm  \\
{\tiny 5} & Joan  & Sep/5 & Therm. & Tymp \\
{\tiny 6} & Helen & Sep/6 & Therm. & Oral \\
{\tiny 7} & Cathy & Sep/6 & Therm. & Oral \\
{\tiny 8} & Cathy & Sep/6 & BPM    & Arm  \\
{\tiny 9} & Susan & Sep/7 & Therm. & Oral \\
{\tiny 10} & Joan  & Sep/7 & Therm. & Oral \\
\cline{2-5}
\end{tabular}
  \vspace{2mm}\caption{} \label{tab:thermometers}
\end{center}
}
\end{minipage}
\end{center}
\vspace*{-6mm}
\end{table}

Table \ref{tab:nurses} contains the name of the nurses in
Tom Waits' ward and the shifts they work in by day. These are the
nurses taking the measurements; since it is a small ward there is
only one nurse per shift with that task. Table \ref{tab:cert}
records the names of the certified nurses in the ward and the year
they got the certification. Table \ref{tab:thermometers} contains the type of
instrument each nurse is using by day (e.g., thermometer, blood
pressure monitor (BPM)), and its type (e.g., arm or wrist for BPM,
and oral or tympanal for thermometer). Each nurse takes all
temperature measurements of the day using the same type of
instrument. This contextual information allows us to assess the
quality of the data in Tables~\ref{tab:temp} and \ref{tab:temp1}.
\boxtheorem
\end{example}
\vspace{-3mm}This paper captures and formalizes the intuition and experience
that data quality is context dependent. This requires an
appropriate formalization of context. In our case, this is given
as a system of integrated data and metadata of which the data
source under quality assessment can be seen as a  particular and
special component.

More precisely, the context for the assessment of a certain
instance $D$ of schema $\mc{S}$ with respect to data quality is given by an
instance $I$ of a possibly different schema $\mc{C}$, which
could be an extension of $\mc{S}$. In other words, $D$ could be
seen as a ``footprint'' of a the contextual, extended database
$I$.

In order to assess the quality of $D$, it has to be ``put in
context'', which is achieved by {\em mapping} $D$ (and $\mc{S}$)
into the contextual schema and data; the extra information in $I$
is what gives context to, and explains, the data in $D$. Actually,
$\mc{C}$ can be more complex than a single schema or instance,
namely a collection of database schemas and instances interrelated
by data- and schema mappings.

The contextual schema and data are not necessarily used to enforce
quality of a given instance. Instead, it can be used to: (a)
Assess the quality of the data in the instance at hand; (b)
Characterize the quality answers to queries; and (c) Possibly
obtain those quality answers to a user query.


Instance $I$ above could be replaced by a much richer contextual
description, e.g. a full-fledged ontology. Along this line, but
still in a classic database scenario, we might define some
additional {\em data quality predicates} on top of $\mc{C}$
\cite{DBLP:conf/er/JiangBM08}. They could be used to assess the
quality of the data in $D$, and also the quality of query answers
from $D$, as we will explore later.

The following contributions can be found in this paper:
\begin{itemize}
\item[(a)] We propose a general model of context and describe how
it can be used for data quality assessment. \item[(b)] We  apply
the context model to:
\begin{itemize}
\item[(1)] Quality (or clean) query answering, i.e. for
characterizing and possibly computing quality query answers. We
concentrate on monotone queries. \item[(2)] Data quality
assessment via some natural {\em data quality measures} that emerge
directly from the model.
\end{itemize}
\item [(c)] We propose algorithms for the previously mentioned
tasks for a few particular, but common, cases. For example, when
we have a contextual instance $I$ that can be used for quality
assessment. We also present an algorithm for quality query
answering under this assumption. Another special case we consider
is when such a contextual instance does not exist. It has to be
first (re)created from the available information and the metadata.
\item [(d)] \ignore{We extend the model by considering {\em external
sources} of data that can be used by the context for data quality
assessment. We also describe criteria and mechanisms for using
them. \item[(f)]} We indicate how our general framework could be
naturally extended in subsequent work to include other features,
like  externally defined quality predicates\ignore{, and {\em
explanations} for the presence of poor quality data. \item[(g)]} \item[(e)] In
addition, we propose {\em a general notion of context}, not only
for data quality purposes. In the light of this general model, we
discuss other tasks that can be undertaken on the basis of
contexts. The contexts used in data quality are then shown to be a
special case of this general and abstract framework.
\end{itemize}

The rest of the paper is organized as follows. In Section
\ref{sec:prel} we introduce a few basic notions of data
management. In Section \ref{sec:contexts} we introduce our
general, abstract notion of context, and we show how it can be applied
in data management in general. In Section
\ref{sec:framework}, we present a general framework for contextual
data quality assessment, introducing intended quality instances and quality query answers. In Sections \ref{sec:perfect} and \ref{sec:noI}, we
consider in more detail two special cases of the general
framework. In Section \ref{sec:qualMeas} we propose some measures for data quality assessment in the presence of multiple quality instances.
\ignore{In Section \ref{sec:external}, we extend the model of
context with the possibility of accessing and using external
sources for data quality analysis.} We discuss related work in Section \ref{sec:disc}. We draw final
conclusions and point out to ongoing and future work in Section
\ref{sec:discussion}. This paper builds on  \cite{birte10}, and extends it in several ways.
In the Appendix we develop some ideas on contexts that access external sources.

\section{Preliminaries} \label{sec:prel}

We will consider relational schemas, say $\mathcal{S}$, with
database predicates $R, \ldots \in \mathcal{S}$, and an underlying
data domain $\mc{U}$. A schema determines a language
$L(\mathcal{S})$ of first-order predicate logic. Queries and views
are defined by formulas of $L(\mathcal{S})$. In this paper we
consider only conjunctive queries and views, i.e. with definitions
of the form
\begin{equation} \label{eq:eq}
\mc{Q}(\bar{x})\!: \ \exists \bar{y}(A_1(\bar{x}_1) \wedge \cdots \wedge A_n(\bar{x}_n)),
\end{equation}
where the $A_i(\bar{x}_i)$ are atoms with database or built-in predicates. Here, $\bar{x}$ contains the free variables of
the query (or view), i.e. that appear in some $\bar{x}_i$, but not in $\bar{y}$. The atoms  may contain domain constants
(i.e. element of $\mc{U}$).
Then, these queries and views
are monotone. Conjunctive queries and unions
thereof can be expressed in non-recursive Datalog with built-ins
\cite{AHV95,CGT90}; i.e by a finite set of rules of the form:
\begin{equation} \label{eq:queryForm}
\nit{Ans}_{\mathcal{Q}}(\bar{x}) \leftarrow R_1(\bar{x}_1), \ldots, R_n(\bar{x}_n), \varphi,
\end{equation}
where $R_i \in \mathcal{S}$, $\bar{x}, \bar{x}_i$ are tuples of variables with $\bar{x} \subseteq \cup_i \bar{x}_i$, and
$\varphi$ is a conjunction of built-in atoms.  $\nit{Ans}_\mathcal{Q}$ is a new predicate
whose extension collects the query answers. Sometimes, we will identify the query with the answer predicate, simply writing
 $\mathcal{Q}(\bar{x}) \leftarrow R_1(\bar{x}_1), \ldots, R_n(\bar{x}_n), \varphi$.

An instance $D$ for schema  $\mc{S}$ is a finite set of
ground atoms (no variables). If $R \in \mc{S}$ is a database predicate,
$R(D)$ denotes its extension in $D$. That is, for each database predicate $R \in \mc{S}$, $R(D) \subseteq D$.
Instances $D, R(D), \ldots$ are those that will be under quality assessment with respect to
to a {\em contextual system}. Similarly, $\mc{Q}(D)$ denotes the set of answers to query $\mc{Q}$ from
instance $D$, and $V(D)$ denotes the extension of the view $V$ on $D$.

For a given schema, say $\mc{S}$, integrity constraints (ICs) are sentences written in $L(\mc{S})$. An instance for
$\mc{S}$ is consistent when $D$ satisfies a given set $\Sigma$ of ICs, denoted $D \models \Sigma$.
For more basic concepts of
relational databases see \cite{AHV95}.

We will assume basics concepts related to {\em schema mappings}, as those found in virtual data integration systems (VDISs) \cite{lenz02,BerBra04},
data exchange \cite{Kol06,arenas10}, or in peer data management systems
\cite{BerBra07}. (See \cite{DeGiac07} for connections between these three areas.)
In general terms, schema mappings take
the form of correspondences between two formulas, like queries or
view definitions, each of them containing predicates from a single
or several schemas. In particular, a data source under
assessment $D$ may have a schema that is mapped into a contextual schema.

We assume that the reader knows the basic concepts of virtual data integration systems, in particular,
the notions of {\em open} (or {\em sound}) source, {\em closed} (or {\em complete}) source, and
{\em exact} (or {\em clopen}) source \cite{lenz02,BerBra04,GM99}.  For summary
and reference, we list below some common forms of associations, or mappings:
\begin{itemize}
\item[1.]
$\forall \bar{x}(S(\bar{x}) \rightarrow
\varphi_{_\mathcal{G}}(\bar{x}))$, where $S$ is a relational
predicate of a data source and $\varphi_{_\mathcal{G}}(\bar{x})$
is a conjunctive query over a global relational schema
$\mathcal{G}$. These association can be found in VDISs under the {\em
local-as-view} (LAV)  paradigm with open (or sound) sources.
\item[2.]  $\forall
\bar{x}(\psi_{_\mathcal{S}}(\bar{x}) \rightarrow G(\bar{x}))$, as
found in {\em global-as-view} (GAV) VDISs with open sources, where
$\psi_{_\mathcal{S}}(\bar{x})$ is a conjunctive query over the
union $\mathcal{S}$ of relational source schemas, and
$G$ is a global relational predicate.
\item[3.] $\forall
\bar{x}(\psi_{_\mathcal{S}}(\bar{x}) \rightarrow
\varphi_{_\mathcal{G}}(\bar{x}))$, as in {\em
global-and-local-as-view} (GLAV) VDISs with open sources. They are
associations between views (or queries) over the source schemas and a global schema, resp.
\end{itemize}
We can see these schema mappings as {\em metadata}, i.e.  data about data; in this case describing data associations. This
 second layer of data can be stored and computationally processed \cite{Arenasetal2010,Bernstein}. Other
 examples of metadata are
relational schemas, integrity constraints (ICs), view
definitions, access and privacy restrictions, trust, and also quality constraints \cite{fan08}.

\section{On Contexts and their Uses}\label{sec:contexts}

As we have stressed above, a full assessment of the quality of
data cannot be conducted in isolation; it only makes sense in a
broader setting. Quality depends, among other things, on the
context that allows us to make sense of the data and assess it. As
expected, the notion of context is not only related to data
quality, but to many other activities, within and outside computer
science.

In computer science we find the term ``context'' in several places , e.g.
databases, semantic web, knowledge representation, mobile applications, etc. It is usually used under {\em context awareness} \cite{Baldauf}, e.g.
 context-aware search, context-aware databases (and query answering), context-aware mobile devices \cite{contextAware}, etc. . However,
most of the time there is {\em no explicit notion of context}, but
only some algorithms that take into account (or into computation)
some obvious contextual aspects. Most typically, time and
geographic location, i.e. particular {\em dimensions}, and not
much beyond.

This makes it clear that there is a lack of research
around the notion of context, at least as used in computer science
and engineering. A precise and formalized general notion of
context becomes necessary. This applies in particular to the use
of contexts -and metadata in general- for data quality assessment
and data cleaning.

It is important to emphasize that, as with contexts in computer
science, there hasn't been much scientific or fundamental research
in the area of data quality and data cleaning. In general,
research in this area tends to be  rather {\em ad hoc}, where
vertical, non-extendable, non-adaptable solutions (usually of an
algorithmic nature) are provided for specific problems and in
specific domains. We are not aware of the existence of an
all-encompassing, general logical theory of contexts, even less
for those that appear in data management. Only some aspects of
them have been used and formalized.

In this work we make an
attempt to contribute with some fundamental research around
certain aspects of general contexts, or better, a general theory
of context that can be applied, in particular, to data management;
and even more specifically, to data quality.
After all, it is clear that the meaning, usability
and quality of data, among other features of data, largely depend
on the context in which data is placed or handled.

Our general perception and formalization of contexts is inspired
by their use in data quality assessment and cleaning. We see
contexts as a form of metadata that can be formalized as a
semantic layer and represented as an ontology, or more generally, as a
{\em theory}.  With this
intuitions, we now describe some of the elements that we envision
in such a general notion of context.

First of all, what is ``put in context" is a logical theory, say
$\mc{T}$. The actual context is another separate logical theory,
say $\frak C$. These two theories are expressed in corresponding
logical languages with their logical semantics. Theories $\mc{T}$
and $\mathfrak{C}$ may share some predicate symbols. The
connection between $\mc{T}$ and $\mathfrak{C}$ is established
through predicates and {\em logical mappings}, i.e. logical
formulas (cf. Figure \ref{fig:genCont}). For example, we could
have mappings of the kind found in virtual data integration or
data exchange (cf. Section \ref{sec:prel}); or in mathematical
logic under ``interpretation between theories" \cite[sec.
2.7]{enderton}, a notion that allows us sometimes to embed a
theory into another.

In principle, theories $\mc{T}$ and $\frak C$
can be written in any formal logic, not necessarily classical
predicate logic, and the same applies to the mappings.

\begin{figure}
\vspace*{-1mm}
    \centering
        \epsfig{file = 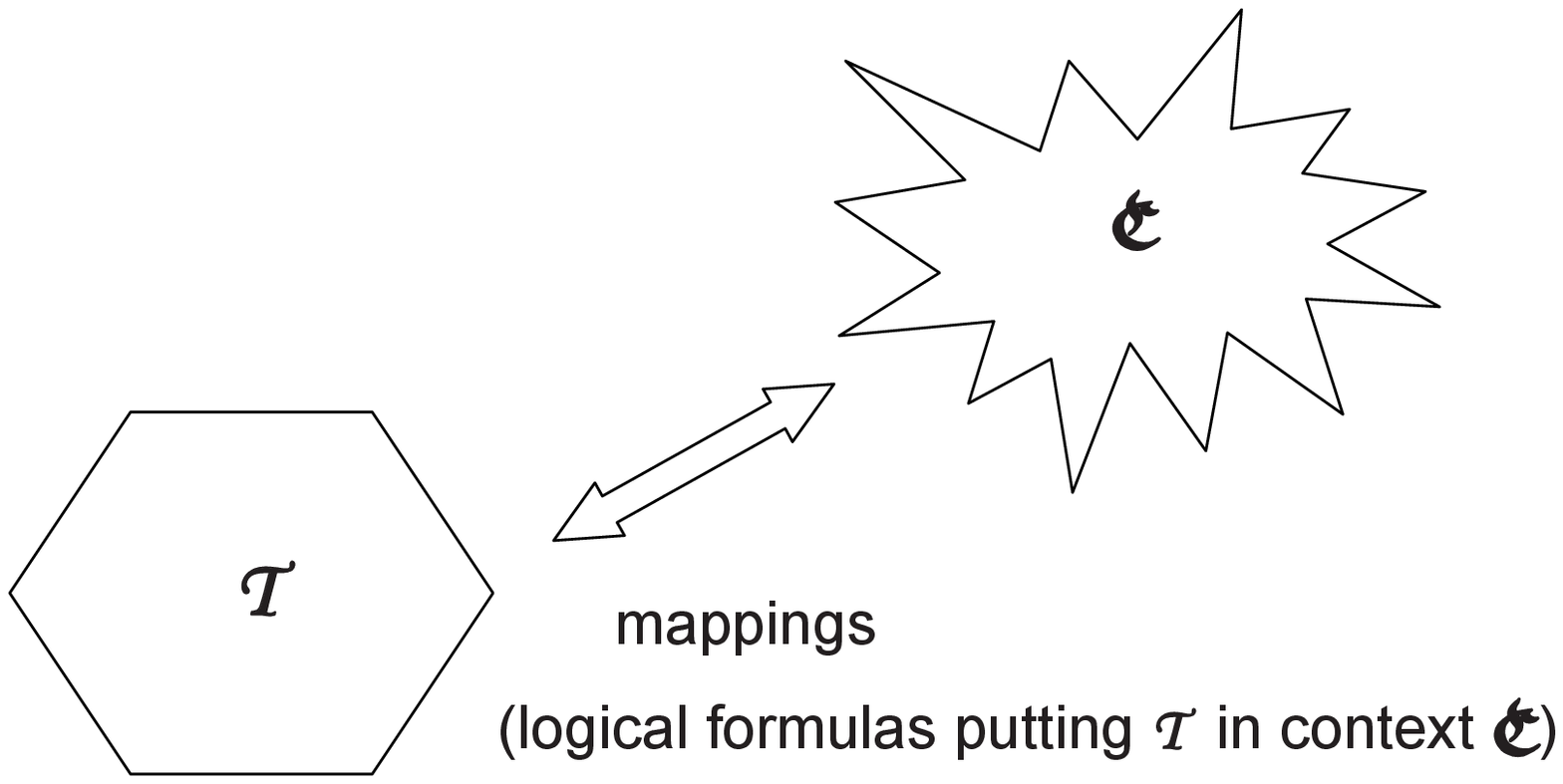,angle=-90,width=80mm}
    \caption{General Contexts}
    \label{fig:genCont}
    \vspace*{-1mm}
\end{figure}

\ignore{
\begin{wrapfigure}{r}{6cm}
\vspace{-6mm}\epsfig{file=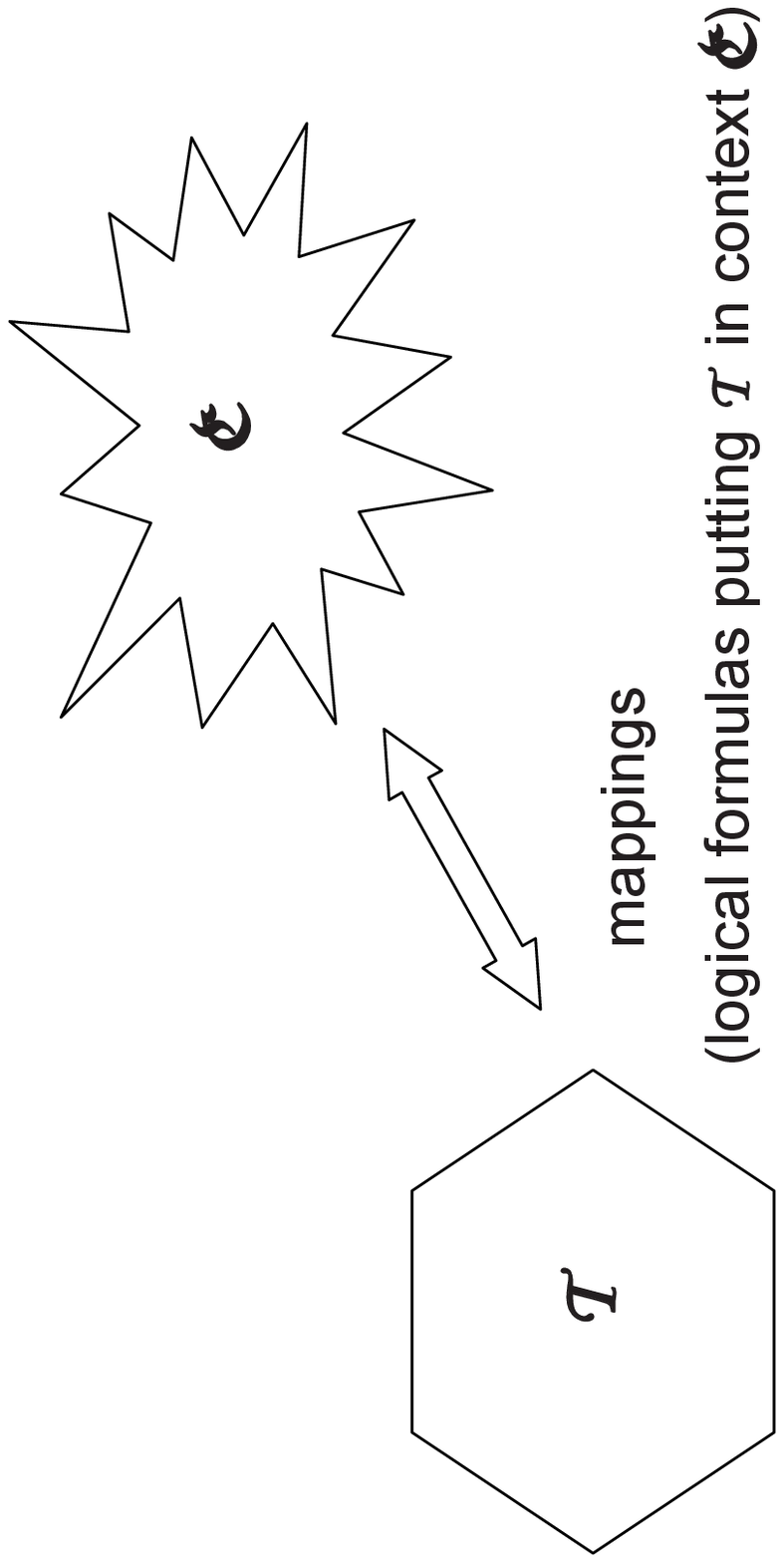,width=6cm}
\caption{General Contexts}
\label{fig:genCont}
\end{wrapfigure}
}

\ignore{
\begin{figure}
\centerline{\epsfig{file=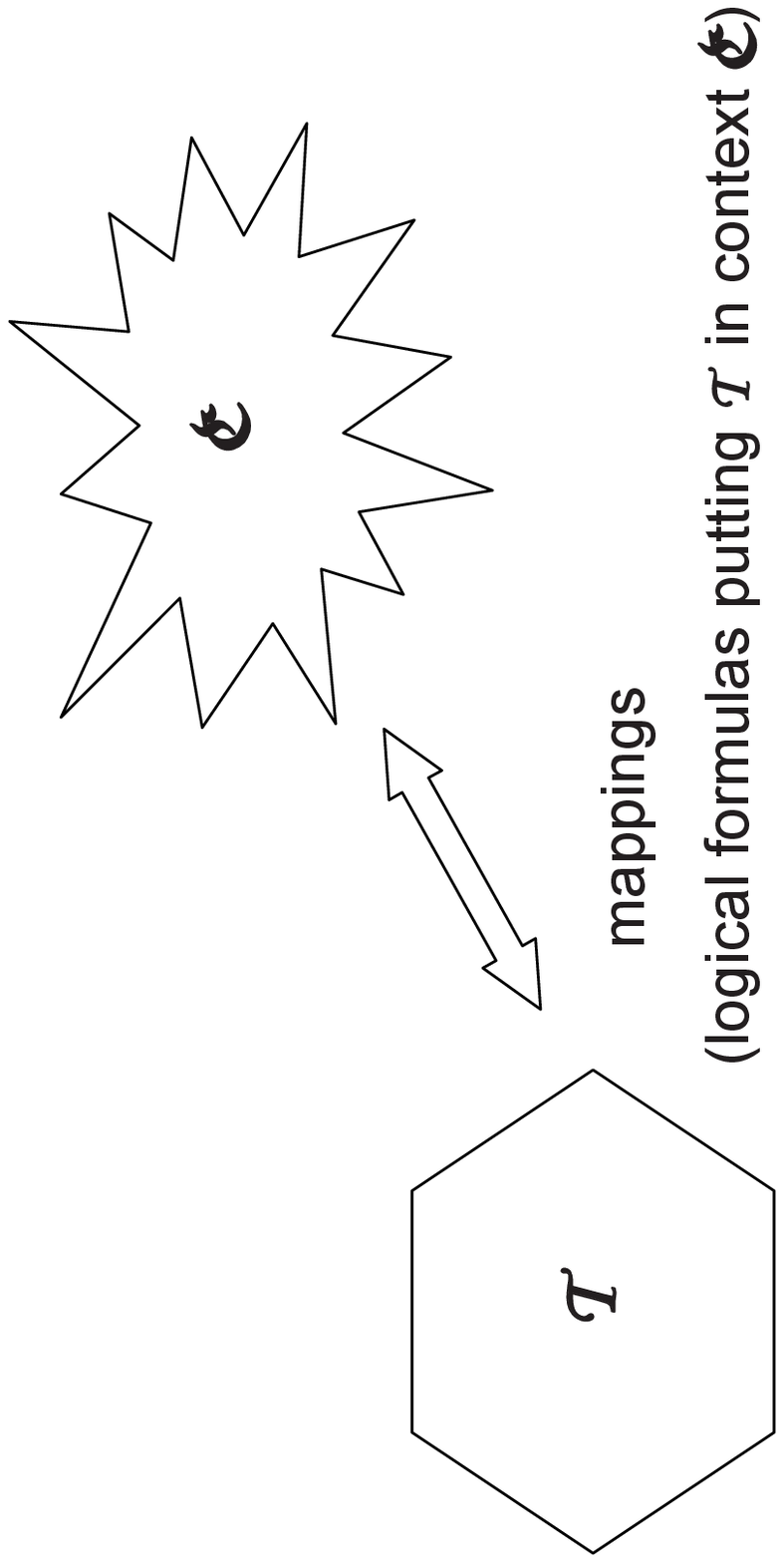,width=6cm}}
\caption{General Contexts}
\label{fig:genCont}
\end{figure}}


Several uses of, and tasks associated to, contexts naturally come
to our mind; among them:
\begin{itemize}
\item[(a)]
Capturing and narrowing down {\em semantics} at $\mc{T}$'s level. This can be achieved by defining in $\frak{C}$
predicates that are used in $\mc{T}$, e.g. ``time close to noon"
in our running example. Context $\frak{C}$ can also contribute
with additional semantic constraints on predicates used in
$\mc{T}$, e.g. ICs on table {\bf TempNoon} in the example. In
particular,  {\em quality constraints} \cite{fan08,bertossiBravo} could be used
as semantic constraints at this level.

\ignore{\comfr{I didn't know what to do with disambiguation. I think it
belongs more to (b) than to (a), but I wasn't sure. It could also
be a separate item.}
}

\item[(b)]
Providing on the basis of $\frak{C}$, a notion and
representation of the {\em sense} of terms in $\mc{T}$. Sense is
intuitively associated to context, and it has been a subject of
preliminary investigation  in data quality
\cite{DBLP:conf/er/JiangBM08}. Moreover, the study of {\em sense
vs. denotation} (or reference) has been a subject of logical
investigation at least since Frege's work \cite{frege1892}. The
notion of sense should be revisited in a contextual framework.
A context should also support {\em disambiguation} of terms appearing in $\mc{T}$. This is also
related to meaning or semantics, and a typical contextual task.

\item[(c)] Providing for {\em dimensions} and {\em points of
view}, to be used for analysis and understanding of $\mc{T}$'s
knowledge. A general definition and formalization of dimension
based on contexts is worth exploring. In particular, to be applied
to problems of data quality, as view points for quality
assessment.

\item[(d)] Specifying and using notions of {\em relevance} for theory $\mc{T}$.

\item[(e)] Providing the conceptual basis and tools for {\em explanation, diagnosis}, and analysis of {\em causality} \cite{suciu10}.

\item[(f)]  Capturing through context $\frak{C}$ {\em common
sense} assumptions and practices to be applied to $\mc{T}$.

\item[(g)] Different forms of  {\em assessment} of $\mc{T}$, e.g. data quality.
\end{itemize}
This is only of partial list of intuitions, notions, and tasks
that we usually associate to the concept of context. Each of them,
and others, could give rise to a full and long term research
program. Some of the open research directions are related to the
identification of representation formalisms, and also
computational processing mechanisms for/of contextual information,
in combination with the theory (or data) it is contextualizing.


It is important to realize that our context-dependent data quality
assessment problem becomes a particular case of our general
concept of context. In the rest of this section, we briefly
provide support for this claim, by indicating how the elements of
the contextual framework in Figure~\ref{fig:genCont} would appear
in the case of data quality assessment. In the following section
we provide specific details.

\subsection{Databases and the general contextual framework} \label{sec:dbGen}

In this section we show in general terms that our abstract model of context
can be applied to data management in general. It may not be obvious that relational databases,
for example, can be seen as theories; and that other data management system, like mediators for virtual data integration, can also be seen
as theories.

In Figure \ref{fig:genCont}, the theory $\mc{T}$ could be a relational database instance
 $D$ for a relational schema $\mc{S}$, with $D$ under quality assessment. Instance
$D$ can be represented
as a theory written in first-order predicate logic (FOPL) by appealing to Reiter's {\em logical reconstruction} of a relational database \cite{reiter84}. It allows to transform the database $D$, usually conceived and treated as a model-theoretic structure, into
a logical theory, $\mc{T}(D)$.

According to Reiter's reconstruction,  query answering from $D$, in particular, becomes expressed as logical entailment and reasoning from $\mc{T}(D)$. Similarly, IC satisfaction becomes logical entailment (of the IC) \cite{ray92}.

For example, relation $\textbf{TempNoon}$ in Example
\ref{exa:intro} (Table \ref{tab:temp1}) can be reconstructed by
means of axiom (\ref{eq:ray}) below plus {\em unique names} and,
possibly, {\em domain closure} axioms \cite{reiter84}.
\begin{eqnarray}
\forall w x y z (\textbf{TempNoon}(w,x,y,z) &~\equiv~& ((w = \mbox{Tom Waits} \wedge \cdots \wedge z =  \mbox{Sep/5}) \label{eq:ray}\\
&&~~~~~~~~~~~~\vee \ \cdots \ \vee \nonumber\\
&&(w =  \mbox{Tom Waits} \wedge \cdots  \wedge z = \mbox{Sep/7}))). \nonumber
\end{eqnarray}
Instance $D$ (think of $\textbf{TempNoon}$) has to be ``put in context", i.e. it has to be mapped into context $\frak{C}$.

Now, $\frak{C}$ could be given as a {\em contextual schema} $\mc{C}$,
plus a possibly incomplete instance $C$ for $\mc{C}$,  and a
set of {\em quality predicates} $\mc{P}$ with definitions in
$\frak{C}$. That is, a context would be, in turn, something like a
a virtual or (semi)materialized data integration system (DIS).

If the contextual data is incomplete, there will
be only a partial specifications of predicates in FOPL.
For example, if the contextual relation $\textbf{S}$ in Table \ref{tab:nurses} is displaying only incomplete data, we will have, in
contrast with axiom  (\ref{eq:ray}) that uses a double implication,  an axiom with a unidirectional implication:
\begin{eqnarray}
\forall x y z (\textbf{S}(x,y,z) &~\leftarrow~& ((x = \mbox{Sep/5} \wedge \cdots \wedge z =  \mbox{Susan}) \ \vee \cdots \ \vee \label{eq:semiray}\\
&&~~~~~~~~~~~~~~~~~~~~~~~(x =  \mbox{Sep/7} \wedge \cdots  \wedge z = \mbox{Joan}))). \nonumber
\end{eqnarray}
In addition,  there will be logical mappings between $D$ and $\mathfrak{C}$, like those of the forms 1.-3.
  at the end of Section \ref{sec:prel}. We could also have ICs and view definitions in $\mathfrak{C}$.
  This is all metadata that can also be expressed as a part of a logical theory.

  To make this more concrete, we could introduce at $\mathfrak{C}$'s level a ``nickname", $\textbf{TN}_\mc{C}$ for relation
  $\textbf{TempNoon}$ in $D$, and have the mapping
  \begin{equation}\forall \bar{x}(\textbf{TempNoon}(\bar{x}) \rightarrow \textbf{TN}_\mc{C}(\bar{x})) \label{eq:cp}
  \end{equation}
  that maps $\textbf{TempNoon}$ into $\textbf{TN}_\mc{C}$. The latter could be further combined with
  $\mathfrak{C}$'s data, to provide the desired information about the nurse certification status of nurse taking temperature, through
  a view (we use Datalog notation for its definition):
  \begin{eqnarray}
  \nit{CertTemp}(m,t,d) &~\leftarrow~& \textbf{TN}_\mc{C}(\text{Tom Waits},m,t,d),\ \textbf{S}(d,s,n), \label{eq:cert}\\
  &&\hspace*{3.3cm}\textbf{C}(n,y),\ \nit{Times}(t,s). \nonumber
  \end{eqnarray}
This view collects the temperature measurements ($m$), with their dates ($d$) and times $(t$), that were taken by certified
nurses. Relation $\nit{Times}$
is used to check if a particular time $t$ falls within a shift $s$. The view extension can be used for further analysis of
$D$'s data.

We could also replace the view defined in  (\ref{eq:cert}) by a more general one, e.g.
\begin{eqnarray}
  \textbf{TempNoon}^\nit{Cf}(p,m,t,d) &~\leftarrow~& \textbf{TN}_\mc{C}(p,m,t,d),\ \textbf{S}(d,s,n), \label{eq:cert2}\\
  &&\hspace*{3.3cm}\textbf{C}(n,y),\ \nit{Times}(t,s), \nonumber
  \end{eqnarray}
that now collects the patient names, possibly other than Tom. The new view predicate, $\textbf{TempNoon}^\nit{Cf}$, does not belong
to the schema containing the
original predicate \textbf{TempNoon}. However, it could be seen as a new version of the latter that now takes into account
the certification of nurses as a particular quality concern.

  We can see that the use of a relational ontology, providing  a context for a relational instance $D$, becomes a particular
  case of the general framework illustrated in Figure \ref{fig:genCont}. In the following sections we will present more
  specific details and concrete examples.

\ignore{\comfr{Me costo entender la subseccion \ref{sec:dbGen}, empieza
bruscamente con algo muy tecnico y no queda muy en claro que trata
de explicar. De hecho me parece que la conclusion del ultimo
parrafo "We can see..." no es nada obvia.}  }

\section{Contexts for Data Quality Assessment}\label{sec:framework}

\subsection{The general approach}

    \begin{figure}
\vspace*{-1mm}
    \centering
  \epsfig{file = 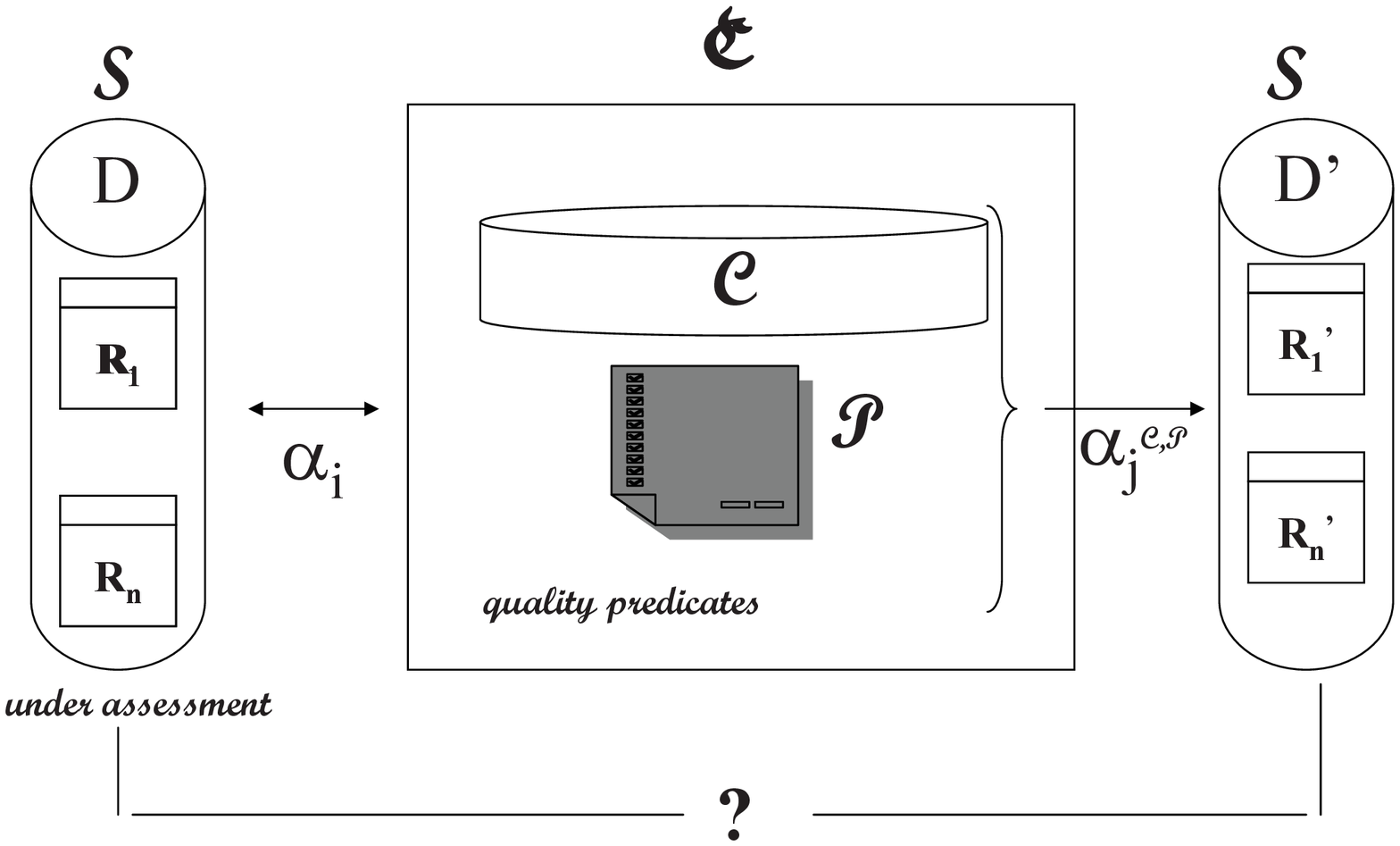,width=80mm}
    \caption{Data Quality Context}
    \label{fig:qualCont}
    \vspace*{2mm}
\end{figure}

We start by describing in general terms our approach to {\em context-based data quality assessment} and {\em context-based quality query answering}. Figure \ref{fig:qualCont} illustrates the main high-level ideas.

Assume we have a database
instance $D$ for schema $\mc{S}$; and  we want to assess $D$'s data quality. $D$ is then put into context $\frak{C}$ via some mappings (the $\alpha_i$ in the figure). The data in $D$'s image in $\frak{C}$ are combined with additional information existing in, or available from, $\frak{C}$. This additional information can be local data at $\frak{C}$,
definitions of quality predicates, additional semantic constraints, and even data from external sources (cf. the Appendix).

\begin{figure}
\vspace*{-1mm}
    \centering
\epsfig{file = 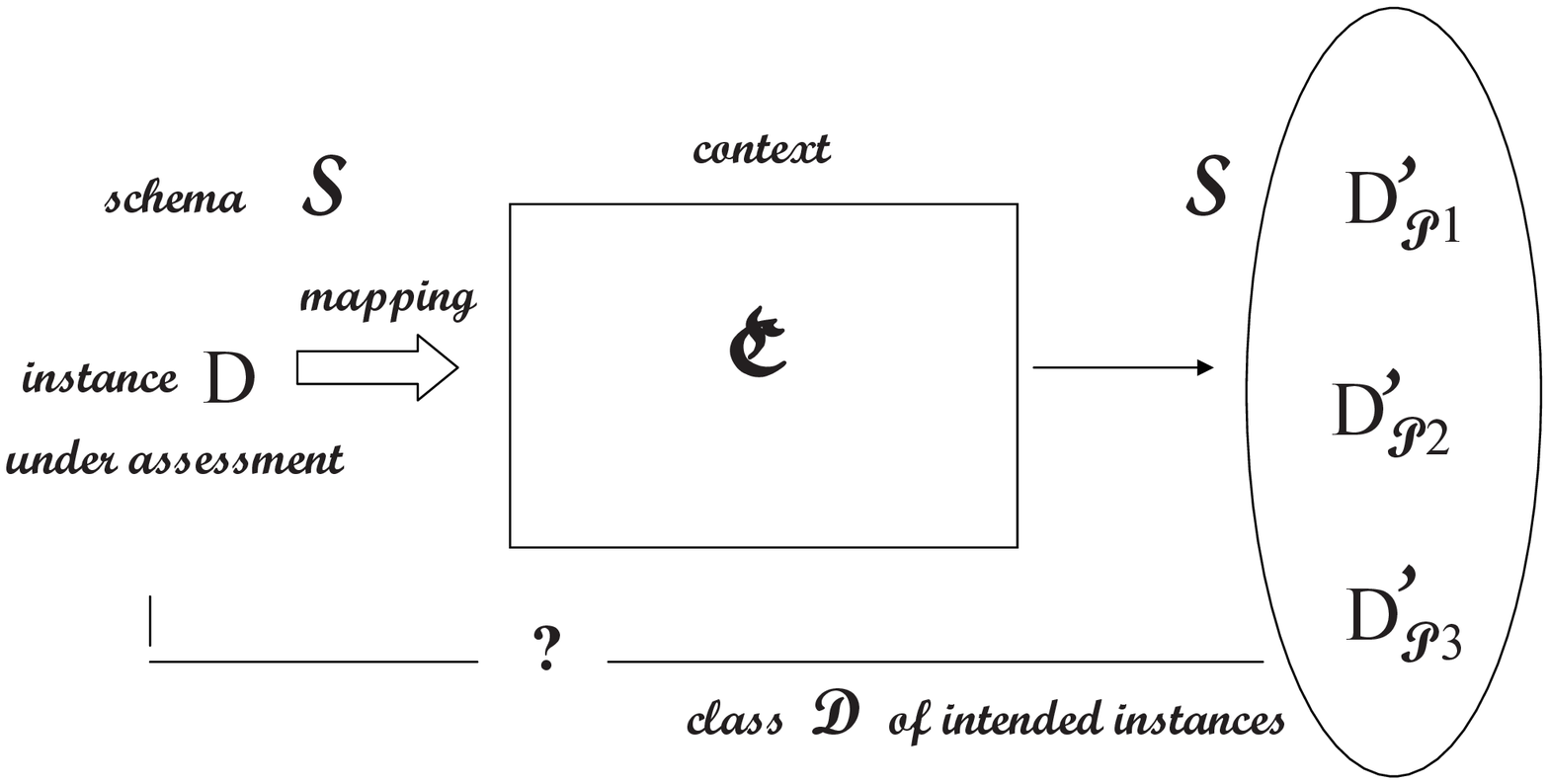,width=95mm}
    \caption{Using a Context for Quality Assessment}
    \label{fig:frameworkPoss}
\end{figure}

The combination at $\frak{C}$ of $D$'s original data with the contextual information produces (via the
mappings $\alpha_j$ on the RHS of the figure) a new version $D'$ of
$D$, or possible a class $\mc{D}$ of several new, admissible  versions of $D$, where the quality concerns are captured or enforced.
This process is also
illustrated in Figure \ref{fig:frameworkPoss}.

  The idea behind context-based data quality assessment is that $D'$  has the correct, clean contents that $D$ should have (e.g. Table \ref{tab:temp1} for $\textbf{TempNoon}$ in Example \ref{exa:intro}).
  And $D$ can be compared with $D'$. For example, the extension for the view in (\ref{eq:cert}) can be compared with
  the subrelation of $\textbf{TempNoon}$ in $D$ that contains the entries for Tom, at least with respect to to the certification status. Similarly, the entire relation  $\textbf{TempNoon}$ could be compared with the relation $\textbf{TempNoon}^\nit{Cf}$ defined in (\ref{eq:cert2}).

As just suggested, the quality of instance $D$ can be measured by comparing it with the resulting instance $D'$ (or collection $\mc{D}$ thereof).
And {\em quality answers} to queries posed to original instance $D$  can be
defined and computed as answers to the same query that are true of $D'$, or of all the admissible $D$' if there are
several of them.  As we will see below, quality assessment and quality query answering are closely related.

The mappings between $D$ (or rather its schema $\mc{S}$) and $\frak{C}$ can take different forms. In this paper,
we will appeal to a common practice in virtual data integration systems of introducing  {\em nicknames} in $\frak{C}$ for the predicates in $\mc{S}$.
This is not strictly necessary, but simplifies the presentation, without losing generality. If an original predicate $R \in \mc{S}$ has a more
elaborate mapping with predicates in in $\frak{C}$, we capture this association as one involving $R'$, the nickname predicate for $R$.

\subsection{The contextual framework}\label{sec:fr}

We have a relational schema $\mc{S} = \{R_1, \ldots, R_n\}$, and also a {\em contextual} relational schema $\mc{C}$ (including built-ins).
The participating schemas are related
by {\em schema mappings}. In particular, the data source under
assessment $D$ may be mapped into the contextual schema.

A common form of mapping is of the form
\begin{equation}
\alpha_{_R}\!: \ \ \forall \bar{x}(R(\bar{x}) \ \rightarrow \
\varphi^{\mathcal{C}}(\bar{x})), \label{eq:footprint}
\end{equation}
where $R \in \mc{S}$, $\varphi^{\mathcal{C}}(\bar{x})$
is a conjunctive query over schema
$\mathcal{C}$.

  We may assume that $\mc{C}$ has a subschema $\mc{S}'$ that is
  a copy of $\mc{S}$ formed by {\em nicknames} $R'$ of predicates $R$ in $\mc{S}$.
  In this case, we may have simple {\em copying mapping} of the form
\begin{equation}
\alpha_{_R}\!: \ \ \forall \bar{x}(R(\bar{x}) \ \rightarrow \
R'(\bar{x})), \label{eq:simple}
\end{equation}
as in (\ref{eq:cp}). We could further
  apply the the closed-world assumption (CWA) \cite{reiter84} to $R'$, obtaining \ $\alpha_{_R}\!: \ \forall \bar{x}( R(\bar{x}) \ \equiv \ R'(\bar{x})$.

   We also have an instance $D$ for
  $\mc{S}$. The extensions of predicates $R \in \mc{S}$ in $D$ are denoted with $R(D)$.
   By applying the mappings $\alpha_i$ to $D$, we obtain (possibly virtual) extensions $R'(D)$ inside $\mathfrak{C}$
  for the nickname predicates $R'$. We obtain an instance
  \begin{equation}
  D' \ := \ \bigcup_{R \in \mc{S}} R'(D), \label{eq:newinst}
  \end{equation} at the contextual level, for
  schema $\mc{S}'$.

  In addition to the contextual schema $\mc{C}$ we may have a set $\mc{P}$ of {\em contextual quality predicates} (CQPs) with
  definitions inside $\mathfrak{C}$. In principle they could  be defined entirely in terms of (or as views over) schema $\mc{C}$.
  However, we keep them separate to emphasize their role in capturing data quality concerns.\footnote{We could also
  consider at the contextual level a set of {\em external predicates} with access to external sources
   that can be used for quality assessment. \ignore{ We will consider this case
in Section \ref{sec:external}.}} We obtain a combined contextual schema $\mc{C} \cup \mc{P}$.
  Typically, each  $P \in \mathcal{P}$ is defined as a
conjunctive view
\begin{equation}
P(\bar{x}) \ \leftarrow \
\gamma^{\mathcal{C}}(\bar{x}), \label{eq:qps}
\end{equation} in terms of elements of
$\mathcal{C}$ (and possibly built-in predicates).

The relations $R'(D)$ inside $\frak{C}$ can be further processed by applying new mappings
$\alpha_j^\mc{C,P}$. Their application logically combine the $R'(D)$s with additional information captured via
the schema (and possibly data for) $\mc{C} \cup \mc{P}$ at $\frak{C}$.

Actually, the ideal, quality
extension of predicate $R \in \mc{S}$ is obtained as the extension of a new predicate (or view) $R'_\mc{P}$.
It is obtained by applying a mapping $\alpha_{_R}^\mc{C,P}$ to $D'$ as in (\ref{eq:newinst}) (possibly only
$R'(D)$) and also additional contextual data and metadata, including CQPs in $\mathcal{P}$.
The mappings or view definitions are as follows:
\begin{equation}
\alpha_{_R}^\mc{C,P}\!: \ \ \forall \bar{x} (\psi^{\mc{C,P}}\!(\bar{x}) \ \rightarrow \ R'_\mc{P}(\bar{x})),\label{eq:filMap}
\end{equation}
where $\psi^{\mc{C,P}}$ is a conjunctive query over schema $\mc{C} \cup \mc{P}$. Notice that $\mc{C}$ contains
schema $\mc{S'}$.
 An example of this kind of definition is (\ref{eq:cert2}).

A special and common case corresponds to definitions via conjunctive views
of the form:
\begin{equation} \label{eq:clean}
\alpha_{_R}^\mc{C,P}\!: \ \ R'_\mc{P}(\bar{x}) \ \longleftarrow \
\varphi_{_R}^\mathcal{C}(\bar{x}_1), \
\varphi_{_R}^\mathcal{P}(\bar{x}_2),
\end{equation}
where $\bar{x} \subseteq \bar{x}_1 \cup \bar{x}_2$, and $\varphi_{_R}^\mathcal{C}(\bar{x}_1),
\varphi_{_R}^\mathcal{P}(\bar{x}_2)$ are in their turn conjunctions
of atomic formulas with predicates in \ca{C}, \ca{P},
respectively.
A particular case is obtained when in (\ref{eq:clean}) there are
no CQPs, i.e.
\begin{equation} \label{eq:noPreds}
R'_\mc{P}(\bar{x}) \ \longleftarrow \ \psi_{_R}^\mathcal{C}(\bar{x}_1). \vspace{-3mm}
\end{equation}
However, we still keep the subscript $\mc{P}$ to establish the difference with $R'$.
Each predicate $R'_\mc{P}$ can be seen as a {\em quality nickname} for predicate $R \in \mc{S}$.

Intuitively, CQPs can be used to express an atomic quality
requirement requested by a data consumer or met by a data
producer. With them we can restrict the admissible values
for certain attributes in tuples, so that only quality
tuples find their way into a quality version of the database.

Although CQPs can be eliminated by unfolding their Datalog
definitions (\ref{eq:qps}), we make them explicit here, for several reasons:
\begin{itemize}
\item[(a)] First, as mentioned above,
to emphasize
their role as predicates capturing quality requirements.
\item[(b)]
They allow us to compare data quality requirements in a more
concrete way. \
For example, it is obvious that the quality
requirement ``temperature values need to be measured by an oral
\emph{or} tympanal thermometer" is less restrictive than
``temperature values need to be measured by an oral thermometer".
\item[(c)] Our approach allows for the consideration  of CQPs that are not defined only in terms
of \ca{C} alone, but also in terms of other external sources\ignore{(cf. Section \ref{sec:external})}, that is, by view definitions of the form
$P(\bar{x}) \leftarrow \gamma^{\mathcal{C}}(\bar{x}),\gamma^{\mathcal{E}}(\bar{x})$, where $\gamma^{\mathcal{E}}$ is a
formula expressed in terms of atoms that can be evaluated at/by external sources.
\end{itemize}

Several different assumptions can be made at this stage, e.g. about the kind of mappings  involved ($\alpha$ or $\alpha^{\mc{C},\mc{P}}$),
  assumptions about them and their sources of data (e.g. openness, closeness, ...), availability or not of an
      initial contextual instance and assumptions about it (e.g. (in)completeness), etc. Some special cases
      will be considered in Sections \ref{sec:perfect} and \ref{sec:noI}. The case of external sources in considered in the Appendix.

Actually, each of the mappings in
Figure~\ref{fig:qualCont} could  be view definitions (or view associations) of any of the LAV, GAV, GLAV
forms described in Section \ref{sec:prel}, with additional assumptions about openness/closedness of the
participating data sources.\footnote{In
virtual data integration it is possible to assign a semantics and
develop query answering algorithms for the case of sources that coexist under different
combinations of
 openness/closure assumptions \cite{GM99,BerBra04}.}

More precisely,  the different data sources, including the original $D$ and any at the context level,
the definitions of the quality
predicates in terms of elements in $\mathcal{C}$,
the schema mappings and view definitions, etc., determine a collection \ca{I} of {\em admissible
contextual
instances} (ACIs) for  the contextual schema \ca{C}, as it is the case in virtual data integration
and peer data exchange systems. Notice that each ACI $I \in \mc{I}$ may also depend on the original instance $D$
if the latter was mapped into the context.

\subsection{Measuring data quality and quality answers}

For a given ACI $I \in \mc{I}$, by applying the mappings (\ref{eq:filMap}), we obtain (possibly virtual)
extensions $R'_\mc{P}(I)$ for the predicates $R \in \mc{S}$. Notice that the collections of predicate extensions $R'_\mc{P}(I)$s
can be seen as a {\em quality  instance} $D'_\mc{P}(I)$ for the original schema $\mc{S}$.

As a consequence, we can assess the quality of $R(D)$ in instance $D$ through its ``distance" to $R'_\mc{P}(I)$;
and the quality of $D$ in terms of an aggregated distance to $D'_\mc{P}(I)$.

Different notions of distance might be used at this point. Just to fix ideas, we can think of using, e.g. the
numerical distance $|R(D)
\bigtriangleup R'_\mc{P}(I)|$  of the symmetric difference between the two $R$-instances. And for the whole instance, e.g.
the {\em quality measure}:
\begin{equation}
\nit{qm}_{\!0}(D) \ := \ \sum_{R \in \mc{S}} |R(D)
\bigtriangleup R'_\mc{P}(I)|. \label{eq:dist0}
\end{equation}
Since we can have several contextual instances $I \in \mc{I}$, we can have a whole class $\mc{D}$ of
instances $D'_\mc{P}(I)$, with $I \in \mc{I}$.  (Actually, we could have none if contextual ICs are imposed, i.e. at $\frak{C}$'s level.)

  This situation is illustrated in Figure \ref{fig:frameworkPoss}. In the case of multiple ``intended" clean instances,
  related to a whole class $\mc{I}$ of ACIs,
  $D$ would have to be compared with a
  whole class of quality  instances
  \begin{equation}
  \mc{D} := \{D'_\mc{P}(I)~|~I \in \mc{I}\};  \label{eq:calD}
  \end{equation} and more elaborate measures of distance
  can be used. We present them in Section \ref{sec:qualMeas}, after developing a scenario, in Section \ref{sec:noI}, where
  those multiple instances naturally appear.

At this point we can introduce the notion of {\em quality
query answering} about (or from) $D$. The idea is to retrieve
clean answers from the original instance $D$. Since the latter may
be dirty, direct, classic query answering from $D$ is not intended.
Instead, clean answers can be obtained from $D'_\mc{P}(I)$ (or the
collection of them).

\begin{example}  (example \ref{exa:intro} continued) Consider a query about
patients and their temperatures around noon on Sep/5:
\begin{equation}
\mc{Q}(p,v)\!: \ \exists t \exists d
(\nit{\textbf{TempNoon}}(p,v,t,d) \, \wedge \, d=\mbox{Sep/5}). \label{eq:que}
\end{equation}
The {\em
 quality answers} to this query posed to Table~\ref{tab:temp} should
be $\langle$\text{Tom Waits}, 38.5$\rangle$, namely the projection
on the first two attributes of tuple 1, but not of tuple 2 because
it does not comply with the quality requirements according to the
contextual tables~\ref{tab:nurses}, \ref{tab:cert}, and \ref{tab:thermometers}.
Notice that if the same query is posed to
Table~\ref{tab:temp1} instead, which contains only quality data
with respect to the quality requirements, we get exactly the same
answer. \boxtheorem
\end{example}

\vspace{-3mm}
More precisely, if a query $\mc{Q} \in L(\mc{S})$ is posed to $D$, but only quality answers are
expected, the query is rewritten into a new query $\mc{Q}'$ in terms of the quality nickname predicates $R_\mc{P}'$, and answered on the basis of their
extensions.

For example, query (\ref{eq:que}) is rewritten into
$$\mc{Q}(p,v)\!: \ \exists t \exists d
(\nit{\textbf{TempNoon}'}_{\!\mc{P}}(p,v,t,d) \, \wedge \, d=\mbox{Sep/5}),$$
with Table~\ref{tab:temp1} showing the intended extension of ${\textbf{TempNoon}'}_{\!\mc{P}}$.

\begin{definition}
The {\em quality answers} to a query $\mathcal{Q}(\bar{x}) \in L(\mathcal{S})$
are  those that are {\em certain}, i.e.
\begin{equation} \label{eq:genClean}
 \nit{QAns}^\mc{C}_D(\mathcal{Q}) = \{\bar{t}~|~ D' \models \mathcal{Q}'[\bar{t}], \mbox{ for all } D' \in \mathcal{D}\},
 \end{equation}
 where $\mc{D}$ is as in (\ref{eq:calD}), and $\mathcal{Q}'$ obtained from $\mathcal{Q}$ by replacing each predicates
 $R \in \mc{S}$ by its quality nickname  $R'_\mc{P}$.\boxtheorem
 \end{definition}
Notice that since quality assessment of $D$ is made by comparison of  the contents of $D$ and the contents of $D'_\mc{P}$,
 it can be seen as a particular case of quality query answering, i.e.
the notion of quality answer could be used to define the quality
of instance $D$: \ For each of the predicates $R \in \mathcal{S}$, we
pose the query \ $\mathcal{Q}_R(\bar{x})\!: \ \nit{Ans}_R(\bar{x}) \leftarrow R(\bar{x})$; and obtain the quality answers
in $\nit{QAns}_D^\mc{C}(\mathcal{Q}_R)$. Thus, $\nit{QAns}_D^\mc{C}(\mathcal{Q}_R)$
becomes an instance for predicate $R$, and can be compared with
$R(D)$. This approach would provide a possibly different quality measure for $D$ than the one in (\ref{eq:dist0}).

In following sections we consider and study some relevant special cases of this
general framework. For each of them, we address: (a) The problem of assessing the quality of the
instance $D$ consisting of the
relations $R_1(D), \ldots, R_n(D)$. This has to do with analyzing  how they differ from
ideal, quality instances for the $R_i$.  (b) The problem of  characterizing
and obtaining quality answers to queries that are expected to be
answered by the instance $D$ that is under assessment.

\section{Towards Quality Assessment:  Contextual Instances} \label{sec:perfect}

A  simple and restricted case of the general framework corresponds to one already illustrated in Sections \ref{sec:intro} and
\ref{sec:dbGen}. It  occurs when we have an
instance $D$ at hand that is under assessment; and there is a material contextual instance $I$, in such a
way that $D$ can be seen as a materialized
view or a {\em footprint} of $I$. Instance $I$ serves  as a reference table
and a basis for the assessment of $D$. Through additional management of $I$ (an indirectly $D$), via quality concerns, it is possible to
obtain an intended, clean version of $D$.

In this situation we can map each relation $R \in \mc{S}$ into the context by means of a definition
of the form   (\ref{eq:simple}), where  $R'$ is a contextual predicate (in $\mc{C}$) that is a nickname for $R$. This is
the case in
(\ref{eq:cp}).

We do not necessarily assume that the mappings (\ref{eq:simple}) are {\em closed}, or, more precisely that $R(D)$ is an {\em exact}
source for $R'$. This is because  $R'$ already has an extension
according to $I$, and there could be a discrepancy between $R'(I)$ and $R(D)$. In consequence, if $R(D)$ is
not assumed to be closed as a source for $R'$, then it holds $R(D) \subseteq R'(I)$.

In order to
recover $R(D)$ as a footprint or materialized view of $R'(I)$, we may have to add additional conditions at $\mathfrak{C}$'s level. For
example, a view definition of the form
\begin{equation}
R(\bar{x}) \ \leftarrow \ R'(\bar{x}), \chi^\mc{C}(\bar{x}'), \label{eq:fp}
\end{equation}
where $\chi^\mc{C}(\bar{x}')$ is a
conjunction of contextual atoms (including built-ins), and $\bar{x} \subseteq \bar{x}'$. However, our main goal is not reobtaining
$D$, but obtaining a ``quality version" of $D$.

In order to do so, we impose additional conditions on the $R'$s, expressed with additional predicates in $\mathfrak{C}$, that can be
 built-in or defined, in particular, those in the set $\mc{P}$ of {\em quality predicates}. We will generically call all those additional
 predicates ``quality predicates", and will be also generically identified with those in $\mc{P}$. As a consequence,
 we obtain for each  predicate $R \in \mc{S}$, an contextual instance $R'_\mathcal{P}(I)$ via a
view definition of the form (\ref{eq:clean}).

Notice that $R'_\mathcal{P}(I)$ can also be seen as an instance for $R$. Actually, from this point of view,
with definitions like those in (\ref{eq:clean}), we can also capture definitions of the form (\ref{eq:fp}),
making
$R'_\mathcal{P}(I)$ coincide with $R(D)$. However,  we are interested in a quality version of
$R(D)$, e.g. $R'_\mathcal{P}(I) \subsetneq R(D)$, with sufficiently strong additional
conditions. In this case, we would be obtaining an
ideal instance for predicate $R$ through  $I$ (that includes the original $D$).

Summarizing,  we obtain an instance for schema \ca{S}:
\begin{equation}
D_\mc{P}'(I) = \{R'_\mc{P}(I)~|~R \in \mathcal{S} \mbox{ and } R'_\mc{P}
\mbox{ is defined by  any of (\ref{eq:filMap})-(\ref{eq:noPreds})}\}.
\end{equation}
As expected, there may be differences between $D$ and
$D_\mc{P}(I)$. The latter is intended to be the
 clean version of $D$.

Since $R' \in \mc{C}$ is expected to
 appear as a $\mc{C}$-atom $R'(\bar{x})$ in any of any of (\ref{eq:filMap})-(\ref{eq:noPreds}),
 it holds $R'_\mc{P}(I) \subseteq R'(I)$ for each $R \in
\mathcal{S}$. Furthermore, if condition $\chi^\mc{C}$ in (\ref{eq:fp}) is included (implied by)
the conditions on the RHS of (\ref{eq:filMap})-(\ref{eq:noPreds}), it will also hold: \
$R'_\mc{P}(I) \subseteq  R(D)$. This would capture the fact that $R'_\mc{P}(I)$ is a further refinement
of $R(D)$ obtained via the contextual information.

\begin{example}  (example \ref{exa:intro} continued) \ \label{ex:cont}
Schema \ca{S} contains $\textbf{TempNoon}(\nit{Patient,}$ $\nit{Value, Time, Date})$, a database predicate, whose
instance
in Table~\ref{tab:temp} is under assessment.

The contextual schema
\ca{C} contains the database predicates $\textbf{S}(\nit{Date, Shift,
Nurse})$, $\textbf{C}(\nit{Name, Year})$, and $\textbf{I}(\nit{Nurse, Date,
Instr, Type})$. For them we have instances:
Tables~\ref{tab:nurses}, \ref{tab:cert}
and~\ref{tab:thermometers}, respectively.

In addition, \ca{C}
contains a predicate $\textbf{M}(\nit{Patient, Value, Time, Date, Instr})$,
which records the values of all measurements performed on patients
by nurses (e.g., temperature, blood pressure, etc.), together with
their time, date, instrument used (e.g., thermometer, blood
pressure monitor). An instance for it is in
Table~\ref{tab:measurements}.

\ignore{
\begin{table}
\vspace*{-4mm}
\begin{center}
\hspace*{-5mm}
\begin{minipage}{47mm}
{\small \setlength{\tabcolsep}{2pt}
\begin{center} } }

\begin{table}
\vspace*{-3mm}
\begin{center}
\hspace*{-8mm}\begin{minipage}{47mm}
{\small \setlength{\tabcolsep}{2pt}
\begin{center}
\begin{tabular}{r|l|r|r|l|l|l|}
\multicolumn{1}{l}{} & \multicolumn{5}{c}{\textbf{M}} \\
\cline{2-6} & \textbf{Patient} & \textbf{Value} & \textbf{Time} &
\textbf{Date} & \textbf{Instr} \\
\cline{2-6}
{\tiny 1}& T. Waits & 37.8 & 11:00 & Sep/5 & Therm. \\
{\tiny 2}& T. Waits & 38.5 & 11:45 & Sep/5 & Therm. \\
{\tiny 3}& T. Waits & 38.2 & 12:10 & Sep/5 & Therm. \\
{\tiny 4}& T. Waits & 110/70 & 11:00 & Sep/6 & BPM  \\
{\tiny 5}& T. Waits & 38.1 & 11:50 & Sep/6 & Therm. \\
{\tiny 6}& T. Waits & 38.0 & 12:15 & Sep/6 & Therm. \\
{\tiny 7}& T. Waits & 37.6 & 10:50 & Sep/7 & Therm. \\
{\tiny 8}& T. Waits & 120/70 & 11:30 & Sep/7 & BPM  \\
{\tiny 9}& T. Waits & 37.9 & 12:15 & Sep/7 & Therm. \\
\cline{2-6}
\end{tabular}
\end{center} \caption{} \label{tab:measurements}
}
\end{minipage}
\end{center}
\vspace*{-3mm}
\end{table}

Relation $\textbf{TempNoon}(\nit{Patient, Value, Time,
Date})$ can be seen as a materialized view of the instance  in Table
\ref{tab:measurements}. It contains, for each patient and
day, only temperature measurements  close to noon.

In this case,
we could this conceive their relationship as established by a mapping of the
form (\ref{eq:footprint}):
$$\forall \bar{x}(\textbf{TempNoon}(\bar{x}) \ \rightarrow \ \exists y \textbf{M}(\bar{x},y),$$
capturing the fact that  $\textbf{M}$ may contain more information that the one
in $\textbf{TempNoon}$ (it is an open mapping).

Or, more specifically for the instance at hand, a  mapping of the form  (\ref{eq:noPreds}), as a
view capturing the temperatures taken between 11:30 and 12:30, with a thermometer:
\begin{eqnarray}
\textbf{TempNoon}_\nit{tm,ins}'(p, v, t, d) &~\leftarrow~& \textbf{M}(p, v, t, d, i), \
\mbox{11:30} \leq t \leq \mbox{12:30},\label{eq:footp}\\&& ~\nit{i}= \mbox{therm}. \nonumber
\end{eqnarray}
The materialization of  this view produces the
instance shown in Table~\ref{tab:temp}, making it a footprint of $\textbf{M}$.

Now, in order to express quality concerns, we can introduce some CQPs.
In this way we will be in position to define the relation that
contains only tuples satisfying the doctor's requirements, i.e.,
that the temperature has to be taken by a certified nurse using an
oral thermometer. In this case, $\mc{P} = \{\nit{Oral(Instr)},$
$\nit{Certified(Patient,Date,Time)},$ and $\nit{Valid(Value)} \}$, whose
elements will be defined in terms of the contextual tables  $\textbf{M},
\textbf{S}, \textbf{C}$ and $\textbf{I}$ (cf. Example \ref{exa:intro}), that are all
part of contextual instance $I$.

In order to facilitate the definitions, we first introduce an
auxiliary predicate,
$\nit{MNT(Patient,Date,Time,Nurse,Instr,Type)}$, that compiles
information about all measurements. It associates each
measurement in $\textbf{M}$ to a nurse and type of instrument used.
\begin{eqnarray}
\nit{MNT(p,d,t,n,i,tp)} &~\leftarrow~& \nit{\textbf{M}(p, v, t, d, i)}, \ \nit{\textbf{S}(d, s, n)}, \ \nit{\textbf{I}(n, d, tp)},\label{eq:nursesAM}\\ &&\mbox{4:00} < \nit{t} \leq \mbox{12:00}, \ \nit{s}=\mbox{morning}. \phantom{ppp}  \nonumber \\
\nit{MNT(p,d,t,n,i,tp)} &~\leftarrow~& \nit{\textbf{M}(p, v, t, d, i)},
\nit{\textbf{S}(d, s, n)},  \nit{\textbf{I}(n, d, tp)},  \label{eq:nursesPM}\\
 & &  \mbox{12:00} < \nit{t} \leq \mbox{20:00}, \ \nit{s}=\mbox{afternoon}. \phantom{ppp}  \nonumber  \\
 \nit{MNT(p,d,t,n,i,tp)} &~\leftarrow~& \nit{\textbf{M}(p, v, t, d, i)},  \nit{\textbf{S}(d, s, n)},  \nit{\textbf{I}(n, d, tp)},  \label{eq:nursesNightPM}\\
 & &  \mbox{20:00} < \nit{t} \leq \mbox{24:00}, \ \nit{s}=\mbox{night}. \phantom{ppp} \nonumber \\
 \nit{MNT(p,d,t,n,i,tp)} &~\leftarrow~& \nit{\textbf{M}(p, v, t, d, i)},  \nit{\textbf{S}(d, s, n)},  \nit{\textbf{I}(n, d, tp)},  \label{eq:nursesNightAM}\\
 & &  \mbox{0:00} < \nit{t} \leq \mbox{4:00}, \ \nit{s}=\mbox{night}. \phantom{ppp} \nonumber
\end{eqnarray}
With the help of this auxiliary predicate, we can define two CQPs:
\begin{eqnarray}
\nit{Oral(p,d,t)} &~\leftarrow~& \nit{MNT(p,d,t,n,i,tp)},
\nit{i}=\mbox{therm},
\nit{tp}=\mbox{oral}. \phantom{ppp}\label{eq:tres} \\
\nit{Certified(p,d,t)} &~\leftarrow~& \nit{MNT(p,d,t,n,i,tp)},
\nit{\textbf{C}}(n,y). \label{eq:uno}
\end{eqnarray}
The first quality predicate is satisfied only when the measurement
(uniquely identified by the patient, the date and the time) was
taken with an oral thermometer (given by the additional conditions
$\nit{i}=\mbox{therm}$ and $\nit{tp}=\mbox{oral}$). The second
predicate can be used to specify that a measurement is made by a
certified nurse.

A third CQP takes care of potential typing errors by checking
that the temperature is in a predefined valid range. It is defined
by:
\begin{equation}
\nit{Valid(v)} ~\leftarrow~ \nit{\textbf{M}(p, v, t, d, i)}, \  \mbox{36} \leq
\nit{v} \leq \mbox{42}.\label{eq:dos}
\end{equation}
With the set $\mc{P}$ of three CQPs in (\ref{eq:tres})-(\ref{eq:dos}), we can define, according to
(\ref{eq:clean}), a new relation: \vspace{-6mm}

\begin{eqnarray}
\textbf{TempNoon}'_\mc{P}(p, v, t, d) &~\leftarrow~&
\nit{M(p, v, t, d, i)}, \ \mbox{11:30} \leq t \leq \mbox{12:30},\label{eq:prime}\\
&& \nit{Valid(v)},
\nit{Oral(p,d,t)},\nit{Certified(p,d,t)}.\vspace{-4mm} \nonumber
\end{eqnarray}
The extension of predicate $\textbf{TempNoon}'_\mc{P}$ is
intended to contain only measurements satisfying the doctor's
requirements. Actually, it  corresponds to the instance shown in
Table~\ref{tab:temp1}. \boxtheorem
\end{example}


\vspace{-3mm}Without considering quality issues, queries are written in the language associated to schema
$\mathcal{S}$, because that is what a user has access to and knows about. If we trust the quality of instance $D$, they would be posed to, and answered from, $D$.
However, if we want to obtain quality answers as determined by the context, the {\em quality answers} to
queries from   $D$ should be, in essence, the answers from its context-dependent quality version
$D_\mc{P}'(I)$ instead.

As a  consequence and a particular case of
(\ref{eq:genClean}), for a query $\mc{Q}(\bar{x}) \in L(\mc{S})$, the set of {\em quality
answers to} \ca{Q} {\em with respect to} $D$ becomes: \vspace{-2mm}
\begin{equation}\label{def:clean}
\nit{QAns}_D^{\mc{C}}(\mathcal{Q}) := {\mathcal Q}'(D_\mc{P}'(I)), \vspace{-2mm}
\end{equation}
where ${\mathcal Q}'$ is obtained from $\mc{Q}$ by replacing each predicate $R$ by its quality nickname $R'_\mc{P}$. If we
see $D_\mc{P}(I)$ directly as an instance for schema $\mc{S}$, outside the context, we pose the original query: \ $\nit{QAns}_D^{\mc{C}}(\mathcal{Q}) = {\mathcal Q}(D_\mc{P}'(I))$.

Since, $D_\mc{P}'(I) \subseteq D$, for monotone queries, e.g. conjunctive queries, it
holds  $\nit{QAns}_D^{\mc{C}}(\mathcal{Q}) \subseteq \mc{Q}(D)$.

In this section we are assuming that the  $R(D)$s are obtained as materialized Datalog views of
the contextual instance $I$ (cf. (\ref{eq:footp}) for an example). As a consequence, clean query answering can be done via view unfolding, when evaluating
the original query on the clean relations $R'_\mc{P}(I)$:

\vspace{2mm}\noindent
{\bf Quality Unfold Algorithm:}  \ (QUA)
\begin{itemize}
\item[1.] Replace each predicate $R$ in \ca{Q} by its corresponding
$R'_\mc{P}$, obtaining query $\mathcal{Q}'$.
\item[2.] Replace $\mathcal{Q}'$
by a query $\mathcal{Q}^\mathcal{C}_\mc{P} \in L(\mathcal{C} \cup
\mathcal{P})$ via view unfolding based on (\ref{eq:clean}).
\item[3.] If
desired, or possible, unfold the definitions of the CQPs,
obtaining the ``quality query'' $\mathcal{Q}^\mathcal{C} \in
L(\mc{C})$, which can be evaluated on $I$.
\end{itemize}
The last step (3.) of the algorithm opens the possibility of
 considering CQPs that are not defined only on top of
schema \ca{C}. This is the case, for example, when they appeal to external sources\ignore{ (cf. Section
\ref{sec:external})},
and also other, lower-level {\em quality predicates} \cite{DBLP:conf/er/JiangBM08}.

\begin{example} (example \ref{ex:cont} continued) \label{exa:query}
Consider the conjunctive query of  $L(\mathcal{S})$ asking about the temperature of the patients
on Sep/5:
\begin{equation}
\mathcal{Q}(\nit{p, v})\!: \ \exists t \exists
d (\nit{\textbf{TempNoon}(p,v,t,d)} \wedge d= \mbox{Sep/5}),
\end{equation}
which in Datalog
notation and using an auxiliary answer-collecting predicate becomes:
\begin{equation} \label{eq:query0}
\mathcal{Q}(\nit{p, v})\!: \ \ \nit{Ans}(p,v)
\ \leftarrow \ \nit{\textbf{TempNoon}(p, v, t, d)}, \ d= \mbox{Sep/5}.
\end{equation}
The first step towards collecting
quality answers is the rewriting of \ca{Q} in terms of the quality-enhanced nickname schema
$\mathcal{S}'$:
\begin{equation*}
\mathcal{Q}'(\nit{p, v})\!: \ \ \nit{Ans}'(p,v) \ \leftarrow \ \nit{\textbf{TempNoon}}'_\mc{P}(\nit{p, v,
t, d)}, \ d= \mbox{Sep/5}.
\end{equation*}
Since $\nit{\textbf{TempNoon}}'_\mc{P}$ is defined by
(\ref{eq:prime}), we can do view unfolding by inserting its definition, obtaining:
\begin{eqnarray}
\hspace*{-3mm}\mc{Q}^\mc{C}_\mc{P}(p, v)\!:  \ \nit{Ans}_\mc{P}'(p,v) &~\leftarrow~&
\nit{\textbf{M}}(p, v, t, d, i), \ \mbox{11:30} \leq t \leq \mbox{12:30}, \label{eq:query}\\
&&\hspace*{-3mm}d = \mbox{Sep/5},
\nit{Valid(v)}, \nit{Oral}(p,d,t), \nit{Certified(p,d,t)}.\nonumber
\end{eqnarray}
This query can be evaluated directly on the contextual instance $I$, which contains relation $\textbf{M}$, by unfolding the definitions (\ref{eq:tres})-(\ref{eq:dos}) of the quality predicates or directly using their  extensions
if they have been materialized.   \boxtheorem
\end{example}

\vspace{-0.8cm}
\subsection{Creating a contextual instance and quality criteria}\label{sec:closed} \vspace{-3mm}

Notice that (\ref{eq:query}) could have quality predicates
that are not defined only in terms of
\ca{C}, but in terms of other external sources or appealing to other filtering criteria. In
this case,  $I$ is not enough, and we may need to
trigger requests for additional, external data.

This independence of the quality predicates from the contextual
data or schema is particularly interesting in the case where we want to
use them to filter tuples from relations in $D$. It also opens the
ground to deal with some  cases where we do not have a given contextual schema (nor
a contextual instance), but only some predicate definitions.

This
situation can be accommodated in the  framework of this section, as follows.
For a predicate $R \in \mc{S}$, we create a copy or {\em
nickname},  $R'$ at the contextual level, obtaining a contextual schema  $\mc{C}$. Each $R'$ shares the arity, the
attributes of $R$, and their domains.

We also have a simple LAV
definition for $R'$: \ $\forall \bar{x}(R(\bar{x}) \leftarrow R'(\bar{x}))$,
considering $R$ as an {\em exact source}, in the terminology of
virtual data integration \cite{lenz02} (this is usual in view
definitions over a single instance). Equivalently, we can define  $R'$
by means of a Datalog rule with its intended application of the CWA on $R'$: \
$R'(\bar{x}) \ \leftarrow \ R(\bar{x})$. In this way we create a contextual
instance $I := \{R'(D)~|~ R \in \mc{S}\}$, for which $D$ is an {\em exact} source.

Instance $I$ can be subject to additional quality requirements as we did earlier in
this section, creating view predicates $R'_\mc{P}$. Their extensions, obtained from
$I$, the quality predicates, and sources mentioned in the latter, can be compared
with the original extensions $R(D)$ for quality assessment of $D$.

In this section we considered the convenient, but not necessarily
frequent, case where the instance $D$ under assessment is a
collection of exact materialized views of a contextual instance
$I$. Alternative and natural cases we have to consider may have
only a partial contextual instance $I^-$ together with its
metadata for contextual reference. We examine this case in Section
\ref{sec:noI}.

\section{Towards Quality Assessment: No Contextual Data} \label{sec:noI}

Against what all the previous examples might suggest, we cannot
always assume that we have a contextual instance $I$ for the contextual schema
$\mc{C}$. There may be {\em some} data for $\mc{C}$, possibly
an incomplete (or empty) instance $I^-$. We might have access to
additional external sources.
Still in a situation like this, data in the instance $D$ under
assessment can be mapped into \ca{C}, for additional composition, analysis under
contextual quality predicates, etc.

In this more general case, a
situation similar to those investigated in virtual data
integration systems naturally emerges. Here, the contextual schema
acts as the mediated, global schema, and instance $D$ as a
materialized data source. In the following we will explore this
connection.

We will consider the case where we do not have a  contextual instance $I$ for schema \ca{C}, i.e. $I^- = \emptyset$.
We could see $D$ as a data
source for a virtual data integration system, $\mathfrak{C}$,
 with a global schema that extends the contextual schema \ca{C}
\cite{lenz02,BerBra04}. We may assume that all the data in $D$ is
related to $\maf{C}$ via \ca{C}, but $\maf{C}$ may have
potentially more data than the one contributed by $D$ and of the
same kind as the one in $D$. In consequence, we assume $D$ to be
an {\em open source} for $\maf{C}$. This assumption will be
captured below through the set of intended or {\em legal} global
instances for $\maf{C}$.

Notice that in Section \ref{sec:closed} we dealt with a similar situation, but there,
the contextual schema basically coincides with the original schema $\mc{S}$, and
$D$ as a source of data for $\mc{C}$ is considered as closed. The case investigated
here could then be seen as
an extension of the one in Section \ref{sec:closed}.

Since not all the data in $D$ may be up to the quality
expectations according to $\mathfrak{C}$, we need to give an account of
the relationship between $D$ and its expected quality version. For
this purpose, as in the previous cases, we extend \ca{C} with a copy $\mc{S}'$ of schema
\ca{S} (or it may already be a part of it): \  $\mc{S}' = \{R'~|~ R \in \mc{S}\}$. All this becomes part of the {\em global schema}
for $\maf{C}$, that may also contain a set $\mc{P}$ of quality predicates. As before, we also include
in the contextual schema the ``quality nickname predicates" $R'_\mc{P}$ for the $R \in \mc{S}$.

\begin{definition} \label{def:legal}
Assume for each $R \in \mc{S}$ we have a the mapping to  schema $\mc{C}$:
\begin{equation}
\forall \bar{x}(R(\bar{x}) \ \rightarrow \ R'(\bar{x})). \label{eq:def}
\end{equation}
A {\em legal contextual instance} (LCI) for system $\maf{C}$ is an instance
$I$ of the global schema $\mc{C}$, such that: \\
(a) For every $R \in
\mc{S}$,  $R(D) \subseteq R'(I)$. Here, $R(D)$ is $R$'s extension in $D$;
 and  with respect to (\ref{eq:def}),
$R(D)$ is seen as an open source; and\\
 (b) \ \ $I \models \forall
\bar{x}(R'_\mc{P}(\bar{x}) \ \equiv \ \varphi_{_R}^\mathcal{C}(\bar{x}) \wedge \ \varphi_{_R}^\mathcal{P}(\bar{x}))$.
\boxtheorem
\end{definition}
This definition basically captures $R$ as an open source under the GAV paradigm. The condition in (a) essentially lifts $D$'s data upwards into
$\maf{C}$. The legal instances have extensions that extend the
data in $D$ when the GAV views in (\ref{eq:def}) defining the $R$s are computed.

The
sentences in (b) act as global integrity constraints (ICs), on
schema $\mc{C}$, and have the effect of (virtually) cleaning the data
 uploaded into $\maf{C}$. The idea is that the nickname $R'$ appears as an atom (condition) in the $\varphi_{_R}^\mathcal{C}$ conjunct
in (b), the nicknames $R'$ obtain data from $D$, and the retrieved data are filtered
with conditions imposed by the quality predicates appearing in  $\varphi_{_R}^\mathcal{P}$
in (b).

We can also consider a variation of the previous case, where, in addition to $D$, we have only a fragment
$I^-$ of the potential contextual instance(s) $I$. That is, we have
{\em incomplete} contextual data. This partial, material fragment has to be taken into account, properly modifying
Definition \ref{def:legal}.

We do so by adding a
condition on $I$: \ (c) \ $I^- \subseteq I$. \ This
requires that the legal
instance $I$ is ``compatible'' with the partial instance $I^-$ at hand.

Notice that if we apply the modified definition, including condition (c), with $I^- = \emptyset$,
we obtain the previous case.\footnote{A partially materialized global instance $I^-$ can be accommodated
in the scenario of VDISs by considering $I^-$ as a separate exact  ``source'' for $\maf{C}$.}

Now, we want to pose queries to $D$, but expecting quality answers. We do so, by posing the same query in terms of
the $R'_\mc{P}$ predicates.
\begin{definition} \label{def:qualAns} A ground tuple $\bar{t}$ is
a {\em quality answer} to
query $\mc{Q}(\bar{x}) \in L(\mc{S})$ iff $\bar{t} \in \bigcap \{\mc{Q}'(I)~|~ I \mbox{ is an LCI}\}$, where $\mc{Q}'$ is
obtained from $\mc{Q}$ by replacing every $R \in \mc{S}$ in it by $R'_\mc{P}$. \boxtheorem
\end{definition}
\vspace{-4mm}As before, we denote with $\nit{QAns}^\mc{C}_D(\mc{Q})$ the set of quality answers to \ca{Q} from $D$ with respect to \ca{C}.
That is, a {\em certain answer semantics} \cite{imiel} is applied to {\em quality query answers}.

\begin{example} \label{ex:quaAns} (example \ref{exa:query} continued)
Let us revisit the query $\mathcal{Q}(\nit{p, v})$ in
(\ref{eq:query}). It is asking about the temperature of the
patients on Sep/5, but with the quality requirements captured by
the context, as we saw in Example \ref{exa:query}.

As before, the instance of $\nit{\textbf{TempNoon}(Patient, Value,
Time, Date)}$ in Table~\ref{tab:temp} is the instance $D$ under
quality assessment. However, in this case we have the contextual
schema, and the quality predicates defined on top of it, but we do
not have full contextual data, only the relations in
Tables~\ref{tab:nurses}, \ref{tab:cert} and~\ref{tab:thermometers}
forming a partial global instance $I^-$. Furthermore, in this
case, we do not even have an instance like the one in Table
\ref{tab:measurements} for predicate $\nit{\textbf{M}(Patient,
Value, Time, Date, Instr)}$ in $\mc{C}$. Basically we have a
possibly partially materialized framework for data quality
analysis and quality query answering.

According to our general approach, we now define a VDIS
$\mathfrak{C}$ with  $D$ as an open source, which will bring data into the context,
producing possibly several LCIs. Each of them
will contain tuples with origin in the given extension of $\textbf{TempNoon}$, but also satisfying the conditions imposed by
(\ref{eq:prime}).
Among them, Table~\ref{tab:temp1} corresponds to the
smallest LCI for $\mathfrak{C}$: No subset of it is an LCI and any
superset satisfying (\ref{eq:prime}) is also an LCI.

 According to Definition~\ref{def:qualAns}, a quality
answer to $\mathcal{Q}(\nit{p, v})$ has to be
obtained {\em from every} LCI for $\mathfrak{C}$. Since
Table~\ref{tab:temp1} is contained in all possible LCI, and we have a monotone query,
it is good enough to pose and answer  the query as usual to/from  Table~\ref{tab:temp1}.

In this way we obtain
the first
tuple in Table~\ref{tab:temp1} as the only answer satisfying the
additional query condition $d= \mbox{Sep/5}$: \
$\nit{QAns}^\mc{C}_D(\mc{Q}) = \{ \langle \mbox{Tom
Waits}, 38.5 \rangle \}$. \boxtheorem
\end{example}
\vspace{-3mm}Since we are considering the original instance $D$ as an open source for the contextual system, we
can take advantage of existing algorithms for the
computation of the certain answers to global queries under the
openness assumption \cite{HalevyVLDB01}. Since we have
conjunctive queries and views, we can use,
e.g. the {\em inverse rules algorithm} \cite{hal00} or extensions
thereof \cite{BerBra04,BraBer03}.

\begin{example} (example \ref{ex:quaAns} continued)
If we invert the definition in
(\ref{eq:footp}) of \linebreak $\nit{\textbf{TempNoon}}$, we get:
\begin{eqnarray}
\hspace*{-1.5mm}\nit{\textbf{M}(p, v, t, d, i)} &~\leftarrow~&
\nit{\textbf{TempNoon}_\nit{tm,ins}'(p, v, t, d)}, \mbox{11:30} \leq t \leq \mbox{12:30},\label{eq:inv}\\
&&\nit{i}= \mbox{therm}. \nonumber
\end{eqnarray}
 We can evaluate
$\mc{Q}^\mc{C}_\mc{P}\nit{(p, v)}$  in (\ref{eq:query}) by unfolding the definition of
predicate \textbf{M} according to (\ref{eq:inv}), and using the fact that $\textbf{TempNoon}_\nit{tm,ins}'$
represents the source $\textbf{TempNoon}$ at the contextual level. We obtain:
\begin{eqnarray}
\hspace*{-3mm}\mc{Q}^\mc{C}_\mc{P}\nit{(p, v)} &~\leftarrow~&
\nit{\textbf{TempNoon}(p, v, t, d)}, \ \mbox{11:30} \leq t \leq \mbox{12:30}, \ \nit{i}= \mbox{therm}, \label{eq:firstRew}\\
&&d = \mbox{Sep/5},\nit{Valid(v)}, \nit{Oral}(p,d,t), \nit{Certified(p,d,t)}. \nonumber
\end{eqnarray}
In their turn, predicates \nit{Valid(v)}, \nit{Oral}(p,d,t), and \nit{Certified} can be unfolded using
(\ref{eq:uno})-(\ref{eq:dos}), transforming query (\ref{eq:firstRew}) into one  in terms of
$\textbf{S}, \textbf{I}, \textbf{C}$,  and also $\textbf{M}$. The latter can be unfolded again using
(\ref{eq:inv}).

The rewritten query can be now evaluated on the instances of
$\nit{\textbf{TempNoon}}$, $\nit{\textbf{S}}$, $\nit{\textbf{C}}$, and $\nit{\textbf{I}}$
(Tables~\ref{tab:temp}, \ref{tab:nurses}, \ref{tab:cert}
and~\ref{tab:thermometers}, respectively). This produces  the same
result as in Example \ref{ex:quaAns}. \boxtheorem
\end{example}

\section{Data Quality Under Multiple Quality Instances} \label{sec:qualMeas}

In  case several possible contextual instances  become admissible candidates to be used
for data quality assessment (as in Section \ref{sec:noI}), some alternatives for this latter task naturally
offer themselves.

First, if we want to assess $D$'s data, we may consider, for
each LCI $I$, the instance $D'_\mc{P}(I) := \{R'_\mc{P}(I)~|~ R \in
\mc{S}\}$, which can also be seen as an instance for schema $\mc{S}$. On this basis, we now introduce two possible {\em data quality measures}.

\begin{enumerate}
\item Given the conjunctive definitions involved, we have  $D'_\mc{P}(I) \subseteq D$.
As a consequence,
$$\nit{qm}_{\!1}(D) \ := \ \frac{|D| -
\nit{max}\{|D'_\mc{P}(I)| : \ I \mbox{ is LCI}\}}{|D|},$$
could be a quality
measure. It is
inspired by the $G_3$ measure in \cite{ma95}. A high quality instance $D$ would have $\nit{qm}_{\!1}(D)$ close to $0$, the value obtained
when $D$ itself is $D$'s only quality instance.

Notice that when there is a single instance LCI $I$ and $D'_\mc{P}(I) \subseteq D$, as here, the quality measure
$\nit{qm}_{\!0}$ in (\ref{eq:dist0}) coincides with $\nit{qm}_{\!1}$.

\item In the same scenario as in 1., a  measure of the {\em degree of quality} of $D$ could be given by
$$r(D) = \frac{|\bigcap_{I  \mbox{ is LCI}} D'_\mc{P}(I)|}{|D|},$$
which is inspired by the Jaccard index \cite{jaccard}. It could be interpreted as the probability of
having a random tuple from $D$ present in all of its quality versions. A value close to $1$ would indicates high
quality data.

\item  Another possible measure is based on quality query answering:  For each predicate $R \in
\mc{S}$, consider the query $\mc{Q}_R\!: R(\bar{x})$; and define
$$\nit{qm}_{\!2}(D) \ := \ \frac{|D \smallsetminus \bigcup_{R \in \mc{S}}\nit{QAns}^\mc{C}_D(\mc{Q}_R)|}{|D|}.$$
The quality answers for each of the queries $\mc{Q}_R$ will produce an instance for predicate $R$ that can be compared
with the initial extension $R(D)$. Again, due to the conjunctive definitions involved, each $\nit{QAns}^\mc{C}_D(\mc{Q}_R)$ is contained in $R(D)$; and then, their union is contained in $D$.
\end{enumerate}
The analysis and comparison of these and other possible quality measures
are left for future work.


\section{Related Work} \label{sec:disc}

Some aspects of contexts have been introduced, used and, sometimes, formalized in the literature, e.g. in
 knowledge representation, data management,  and some other areas where ``context-aware" application, devices,
 and mechanisms are proposed.

 In \cite{mccarthy93}, in traditional knowledge representation, we find a logical treatment of contexts. They are not defined, but {\em denoted} at the object level. In this way a theory specifies their properties, dynamics and combinations. It becomes possible to talk about things holding in certain (named) contexts. To the best of our knowledge, this framework has not been exploited in data management.

 There has been some work on the formalization and use of contexts
done by the knowledge representation community. There are approaches to contexts there that also address some of
our concerns, most prominently, the idea of integration and interoperability of
models and theories.  In \cite{giunch94}, the emphasis is placed on
the proper interaction of different logical environments.

In a similar direction, {\em multi-contexts systems} have also been investigated, and the problem of {\em bridging} them, e.g. using logic programs, is matter of recent and active ongoing research. See \cite{brewka11} for a survey and additional references. Central elements are the {\em bridge rules} between denoted contexts, where each of the latter can have a knowledge base or
ontology of its own. The bridge rules are expressed as propositional logic programming rules. The application of this kind
of multi-context systems in data management and their use for expressing the kind of rich mappings found there are still to be investigated.

 Not necessarily explicitly referring to
contexts, there is also recent work on the integration of
ontologies and distributed description logics
\cite{serafini10}. More specifically around ontologies and the  semantic web, we find additional
 relevant work in this direction.

The
{\em local model semantics} (LMS) \cite{ghidini01,giunchiglia93} is a general semantic framework for contextual reasoning. It is
based on two principles: \ (a) {\em Locality}: Reasoning only uses part of the available knowledge about the world. The portion of knowledge being used is called a context. \ (b) {\em Compatibility}: There are relations between the kinds of reasoning performed in different contexts. These relations are called compatibility relations. \
In this framework different propositional languages are exploited to describe facts in different contexts. The notion of local model, local satisfiability and logical consequence are relative to those languages.
The formalization of contextual reasoning as captured by LMS is illustrated with examples of reasoning with viewpoints and beliefs.

The emphasis is placed on capturing {\em locality} properties of contexts, in particular,
 {\em viewpoints} and {\em dimensions}. They  provide perspectives from where a representation can be seen and analyzed. This captures an important intuition of {\em locality} behind contexts, but not fully the features of {\em generality, extension}
and {\em embedding} that we have proposed and used in our work. In \cite{giunchiglia93,ghidini01}, contexts are models (or sets thereof) rather than theories. The problem of combining  contexts is considered.
In \cite{ghidini98} ideas from (an earlier version of) \cite{ghidini01} are applied to information integration.

The use of contexts in data management has been proposed before (cf.
\cite{bolchini.sigmod-rec07.context-models} for a survey). As expected, there are different ways to
capture, represent and use
contexts. In  data management, we usually find an {\em implicit} notion of context,
 in the form of {\em context awareness}, and commonly associated to
 the {\em dimensions} of data, which are usually time, user and location.

In \cite{tancaCMT,bolchini.acm-trans09.context-data} a context-driven approach is presented for extracting relational data views according to implicit
 dimension or {\em context elements}, like the time, situation and interest of the user. The set of views is built on the basis of a specification of some context dimensions and their current values. This approach introduces a context model
 as a dimension tree, as an array of ambient dimensions. In the {\em context dimension tree} (CDT), the root's children are the context dimensions which capture the perspectives from which the data can be viewed. A dimension value can be further refined with respect to different viewpoints, generating a subtree in its turn. Context elements are essentially attribute-value
pairs, e.g. role=`agent', situation=`on site', time=`today'. Furthermore,
certain constraints can also be specified on a CDT, e.g. when
role is 'manager', the situation cannot be 'on site'.

By specifying values for dimensions, a point in the multidimensional space representing the possible contexts is identified. Accordingly a context or a chunk configuration is a set of dimension values. After such a chunk configuration has been specified, the next task is to associate the chunk configuration with the definition of the corresponding schema. The designer specifies a chunk configuration to tailor the data portion relevant to the corresponding context from the actual data set.

A context specification allows one to select, from a potentially
large database schema, a small portion (a view) that is deemed
relevant in that context \cite{bolchini.er07.view-composition}. Given a context specification, as in
\cite{tancaCMT,carve}, the
design of a context-aware view may be carried out manually or
semi-automatically by composing partial views corresponding to
individual elements in that context.

In some sense, the main purpose in
\cite{tancaCMT} is size reduction \cite{tanca07}, i.e. the
separation of useful data from noise in a given context. This is still in the spirit of capturing
{\em locality} and {\em viewpoints}.  In our work,
however, the main purpose is the embedding into a {\em more general setting}, for quality-based data selection, i.e.
of a subset of data that best meets certain quality
requirements.

 In \cite{spyratos07} an interesting formalization of contexts as applied
 to conceptual modeling is presented. Contexts are sets of named objects, not theories.
 Each object has a set of names and possibly a reference. The reference of the object is another context that stores detailed information about the object. The contents can also be structured through the traditional abstraction mechanisms, i.e. classification, generalization, and attribution. We can say that this work
 captures some intuitions of
 abstraction, generalization and reference.
The interaction between contextualization and traditional abstraction mechanisms is studied, and also the constraints that govern such interactions. Finally, they present a theory for contextualized information bases. The theory includes a set of validity constraints, a model theory, and a sound and complete set of inference rules.

 In \cite{davide09,davide10,davideVLDBJ} contexts are explicit, but undefined, and correspond in essence to dimensions as
 found in data warehouses and multidimensional databases \cite{hurtado05}. When queries are posed to a database via a context, the former is expanded via dimensional navigation and aggregation as provide by the latter, to return more
 informative answers to the query.

Data quality and data cleaning encompass many issues and problems \cite{batini06.data-quality,floris}.
However, not much research can be identified on the use of formalized contexts for data quality assessment
and  cleaning.

The study on data quality spans from the characterization of types
of errors in data \cite{beyond}, through the modeling of
processes in which these errors may be introduced
\cite{modeling}, to the development of techniques for error
detection and repairing \cite{Bleiholder2008}. Most of
these approaches, however, are based on the implicit assumption
that data errors occur exclusively at the syntactic/symbolic
level, i.e. as discrepancies between data values (e.g. Kelvin
vs. Kelvn when referring to temperature degrees).

As argued in \cite{DBLP:conf/er/JiangBM08}, data quality problems
may also occur at the semantic level, i.e. as discrepancies
between the meanings attached to these data values. More
specifically, according to \cite{DBLP:conf/er/JiangBM08}, a data
quality problem may arise when there is a mismatch between the
\emph{intended} meaning (according to its producer) and
interpreted meaning (according to its consumer) of a data value. A
mismatch is often caused by ambiguous communication between the
data producer and consumer; and such ambiguity is inevitable if some
sources of variability, e.g. the type of thermometer used and the
conditions of a patient, are not explicitly captured in the data
(or metadata). Of course, whether or not such ambiguity is
considered a data quality problem depends on the purpose for which
the data is used.

In \cite{DBLP:conf/er/JiangBM08} a framework was proposed for
\emph{defining} both syntactic- and semantic-level data quality in
an uniform way, on the basis of the fundamental notion of signs
(values) and senses (meanings). A number of ``macro-level"
quality predicates are also introduced, based on the comparison
of symbols and their senses (exact match, partial match and
mismatch).

Relevant work on  doing quality assessment of query answering is presented in \cite{Naumann02}.
That work is based on a universal relation \cite{DBLP:journals/tods/MaierUV84}
constructed from the global relational schema for integrating
autonomous data sources. Queries are a set of attributes from the
universal relation with possible value conditions over the
attributes. To map a query to source views, user queries are
translated to queries against the global relational schema.
Several quality criteria are defined in \cite{Naumann02} to qualify the sources,
such as believability, objectivity, reputation and verifiability,
among others. These criteria are then used to define a quality
model for query plans.

According to \cite{Naumann02}, the quality of a query plan is
determined as follows. Each source receives information quality
(IQ) scores for each criterion considered to be relevant. They are then
combined into an IQ-vector where each component corresponds to a
different criterion.

Users can specify their preferences for the
selected criteria, by assigning weights to the components of the
IQ-vector, hence obtaining a weighting vector. The latter is used in its
turn by multi-attribute decision-making (MADM)
methods for ranking the data sources participating in the
universal relation. These methods range from the simple scaling
and summing of the scores (SAW) to complex formulas based on
concordance and discordance matrices.

The quality model proposed in \cite{Naumann02} is
independent of the MADM method chosen, as long as it supports user
weighting and IQ-scores. Given IQ-vectors of sources, the goal is
to obtain the IQ-vector of a plan containing the sources. Plans
are described as trees of joins between the sources: leaves are
sources whereas inner nodes are joins. IQ-scores are computed for
each inner node bottom-up and the overall quality of the plan is
given by the IQ-score of the root of the tree.

\section{Discussion and Conclusions} \label{sec:discussion}





We have proposed a general framework for the
assessment of a database instance in terms of quality properties. The assessment is based
on the comparison with a class of alternative intended instances
that are obtained by interaction of the original data with additional contextual data
or metadata. Quality
answers to a query also become relative to those alternative
instances.

Our framework and above mentioned interaction involves mappings between database
schemas, like those found in data exchange, virtual data
integration, and peer data management systems (PDMSs).

In this work we have undertaken the  first steps in the direction of capturing data quality
assessment and quality query answering as context dependent activities. We
examined a few natural cases of the general framework. We also
made and investigated some assumptions about the mappings, views and queries
involved.

Crucial contextual elements for data quality assessment are the {\em quality
predicates}. They can be quite general, and can even be defined in terms of
{\em external sources} that can be invoked for quality assessment and quality query answering.
This situation arises when the context does not have all the elements to support data quality assessment or quality query answering.
Instead,  the context has access to external sources of data, which are independent from the context and from each other,
may have completely different schemata, and most importantly, may have restrictions on the queries they accept and the answers they provide.
Depending on these restrictions, queries and requests by the context have to be adjusted accordingly. In the Appendix we sketch some of the issues,
problems and solutions. This is still part of our ongoing research.

In this sense, we see our work as a next step after the use of low-level quality predicates
as those proposed in \cite{DBLP:conf/er/JiangBM08}. They capture  more intrinsic quality properties, like data value, syntax, correctness, sense, meaning, timeliness, completeness, etc.
Properties of this kind can (and should at some point)
be integrated in our contextual framework. This is matter of ongoing research. In this work we have proposed a
methodology for capturing and comparing higher, semantic-level data
quality requirements, using context relations and quality
predicates, and we have showed how they are used in query answering.

Our work is quite general and abstract enough to accommodate different forms of
data quality assessment as based on contextual information. We think this kind of work
is necessary due to  the lack of fundamental research around data quality in general.
Actually,
most of the research in the area turns around specific problems and applications, mostly vertical,
that cannot be easily adapted for other problems, scenarios, and application domains. It is necessary
to identify, conceptualize, and investigate  the main general principles and methodologies that underly data
quality assessment and data cleaning.

Our proposed framework should be extended in order to include  more intricate
mappings. More algorithms have to be
developed and investigated, both for quality assessment and for
quality query answering. Much research is still to be done in this area.

Among other prominent problems for ongoing and future research we find a detailed and comparative analysis of the quality measures introduced in this paper,  and also others that become natural and possible. Also the development of (hopefully) efficient mechanisms for computing them is an open problem.

Going beyond relational contexts, we are also considering contexts for data quality assessment that are provided by richer ontologies, e.g.  expressed in semantic web languages, such as DL or OWL. Such contexts would be more general, and
naturally  admit several models in comparison with  the a relational case. Reasoning also becomes a new issue.
In this ontological-contextual direction,  we can make a broader use of the rich logical language for defining quality predicates; and possibly
 also additional information  obtained through experience in data cleaning
and domain knowledge. It would become possible to pose queries that {\em explicitly} ask for
answers that satisfy some quality conditions. Querying databases through ontologies has been the subject of recent
research \cite{poggi08,caliLics,kontchakov10}.

Actually, we have made progress in this direction by introducing both dimensions and ontologies into data quality assessment. In fact, as discussed in Sections \ref{sec:intro} and \ref{sec:disc},
contexts are expected to
provide and support the notions of {\em dimension} and {\em point of view}. Dimensions are natural components of contexts. In \cite{mostafa}
our model
of context for data quality assessment is extended with dimensional elements,  making it possible to do multidimensional assessment of data
quality.
The extended model includes the Hurtado-Mendelzon model of multidimensional databases \cite{hurtado05}. In this way, dimensions can be
properly integrated with the rest of the contextual information. Notice that this approach
 leads to a generalization of the notion
of dimension as a contextual element, going beyond the typical cases found in the literature, such as
geographic and temporal dimensions.

More specifically, in \cite{mostafa} multidimensional contexts are represented
as ontologies written in Datalog$\pm$ \cite{caliLics,cali}. This language is used for
representing dimensional constraints and
dimensional rules, and
also for doing
query answering
based on dimensional navigation,
which -as we have shown in this work- becomes an important auxiliary activity in the assessment
of data.

Our approach to quality versions of a given database as determined by contexts is reminiscent of database
repairs and consistent query answering \cite{bertossi11}. Actually, the latter scenario could be  obtained by means
of a context $\mc{C}$ that has the same relational schema as a given instance $D$ under data quality assessment.
$\mc{C}$ does not have an instance (as in Section \ref{sec:noI}), but does have integrity constraints $\Sigma$ that may not be satisfied by $D$.
Mapping $D$ into $\mc{C}$, and making the mapped version respect $\Sigma$ produces the repairs of $D$ with respect to
$\Sigma$ as the contextual quality versions of $D$. They are consistent instances, i.e. they satisfy $\Sigma$, and minimally
differ from (the mapped version of) $D$. The quality answers to a query become the consistent answers to the query, i.e. those
that can be obtained from all repairs.

Furthermore, the quality measures considered in Section \ref{sec:qualMeas} (and possible others) could be used to measure the {\em degree of
consistency} of $D$ with respect to $\Sigma$. This is subject that deserves much more investigation.

\vspace{2mm}
\noindent{\bf Acknowledgments:} Research funded by the NSERC Strategic Network on BI (BIN, ADC05 \& ADC02) and NSERC/IBM CRDPJ/371084-2008. 



\section{Appendix: \ Contexts with External Sources}
\label{sec:external}

\begin{figure}
\vspace*{-1mm}
    \centering
 \epsfig{file = 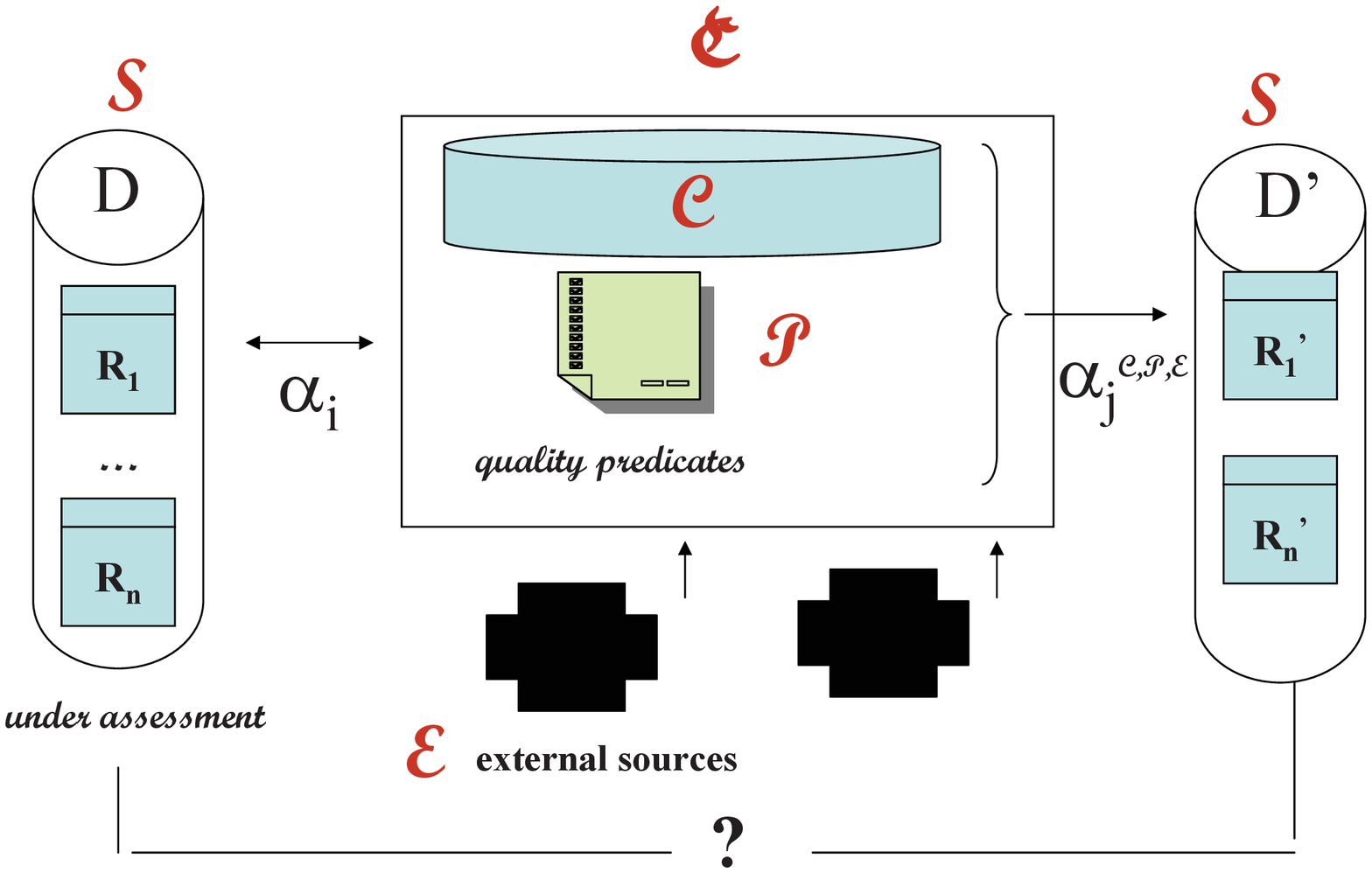,width=90mm}
    \caption{Contexts with External Sources}
    \label{fig:qualContExt}
    \vspace*{2mm}
\end{figure}

As announced before, a context may have access to external sources, as illustrated in
Figure~\ref{fig:qualContExt}. As before,  the
relations $R_i$ in $D$ are under quality assessment
via the contextual system $\frak{C}$ and, in particular, a contextual schema $\mathcal{C}$.
The latter has relational
predicates $C_1, \dots, C_m$, possibly including nicknames $R_i'$s for the $R_i$s.

Now the {\em contextual quality predicates} (CQPs) $P_1, \ldots, P_k$
in set $\mathcal{P}$ can be defined as views in terms of predicates in $\mathcal{C}$ (or others defined purely in
terms of $\mathcal{C}$) and, possibly
{\em external predicates} $E_1, \ldots, E_j \in \mc{E}$ that can be evaluated
only on sources outside context $\frak{C}$.

In some cases, the
combination of schemas  $\mathcal{C}, \mathcal{P}, \mathcal{E}$
can be seen as a single, extended contextual schema. In other
cases, it may be useful to tell them apart. Actually, the external
predicates $E_i$ could be seen as a part of $\mc{C}$, but there is no material data for them
in $\frak{C}$. Alternatively, we could introduce ``nicknames" $E_i'$ in $\mathcal{C}$,
with simple view definitions as mappings, of the form $\forall
\bar{x}(E'(\bar{x}) \equiv E(\bar{x}))$ (or, in Datalog notation,
$E'(\bar{x}) \leftarrow E(\bar{x})$), where $E$ is the predicate at the source. The important issue
in this scenario is that the sources $E_i$ can be accessed upon request, more precisely at (quality) query answering
time.

More precisely, and as before, in order to
obtain quality answers to a query $\mc{Q}$ of the form  (\ref{eq:queryForm}) posed to the instance under
assessment $D$, we
produce instead a query of the form
 (\ref{eq:clean}). However, now the conjunctions $\varphi_{_R}^\mathcal{C}(\bar{x})$ and $\varphi_{_R}^\mathcal{P}(\bar{x})$
may both contain external predicates, associated to external data sources.

The data to evaluate the query will come from the contextual instance $I$ (that is possibly partial and
also partially determined by $D$) and the
external sources.  If the data for the latter are missing, the query evaluation process should trigger {\em ad hoc}
requests for external data. The key issue here is at what point these external predicates are
actually invoked and how.

 If the request for external data is
performed without a value selection, then the external predicate
might possibly return an unnecessarily large instance (of which only a portion
would be used at the contextual level). Or the external source could return no
data if it is not queried with the right access bindings (cf.
below), a problem that has been investigated in data management \cite{CM08,DLN07}.

We now describe a way to address both problems that is based on
properly pushing down value selections. That is, the evaluation can
invoke the external predicates with some bounded variables,
getting in return a much smaller answer set, namely the one
satisfying the value selection. To do this, we make use of
\emph{magic sets techniques} (MST)~\cite{BR87}, a family of bottom-up
query evaluation methodologies that simulate the pushing-down of selections found in
top-down approaches (cf. also \cite{CGT90,AHV95} for more details).

\begin{example} (example \ref{exa:query} continued) \label{exa:queryexternal}
To simplify the presentation, we will use instead of predicate $\nit{\textbf{TempNoon'}}_{\!\!\!\mc{P}}$ defined in (\ref{eq:prime}), a new (clean) alternative
predicate, namely $\nit{\textbf{TempNoon''}}_{\!\!\!\mc{P}}$, which is
defined using only the $\nit{Certified}$ CQP:
\begin{eqnarray}
\textbf{TempNoon''}_{\!\mc{P}}(p, v, t, d) &~\leftarrow~& \nit{\textbf{M}(p, v, t,
d, i)}, \ \mbox{11:30} \leq  t \leq \mbox{12:30},\label{eq:prime2} \\
&&\nit{Certified(p,d,t)}. \nonumber
\end{eqnarray}
Furthermore, assume that $\nit{Certified(p,d,t)}$ is now defined in terms of an external source
$\nit{\#\!C(Nurse, Year)}$, and not as in (\ref{eq:uno}) in
terms of
the local source $\nit{\textbf{C}(Nurse,Year)}$:
\begin{equation}
\nit{Certified(p,d,t)} \leftarrow \nit{MNT(p,d,t,n,i,tp)},
\nit{\#\!C}(n,y). \label{eq:uno2} 
\end{equation}
Predicate $\nit{MNT}$ was defined via (\ref{eq:nursesAM})-(\ref{eq:nursesNightAM}) in Example \ref{ex:cont}. However, for the sake of illustration, we will assume in this example
that it is an extensional (non-defined) predicate.

The external source $\nit{\#\!C(Nurse, Year)}$ contains information
about certified nurses; and $\nit{\#\!C(Nurse, Year)}$ returns the year
of certification if the input nurse appears in the source, and
the constant $\nit{null}$, otherwise. Furthermore, this source only returns an answer (the certification year)
if it is {\em asked about a specific nurse at a time}. That is, the access to this source is subject to {\em binding restrictions}. It is not possible to obtain answers to queries about a whole set of unnamed  nurses. \boxtheorem
\end{example}

The first issue is, in general, how to guarantee that the binding
restrictions are satisfied, i.e., to make sure that input
attributes are always bound at the moment of requesting external
data.

We address that issue by considering only \emph{input guarded rules}: \ for every
rule $r$ (in a Datalog definition) containing an external
predicate $\#E$ in its body, every {\em input variable} of $\#E$
(i.e. that requires a concrete instantiation in a query, like
$\nit{Nurse}$ in the example) has to appear in some previous atom
in the body of $r$, but not as input variable for another external
predicate. This condition can be used to guarantee that input
variables in external predicates can be  bound when the request
for data is made.

\begin{figure}
\vspace*{-1mm}
    \centering
       \epsfig{file = 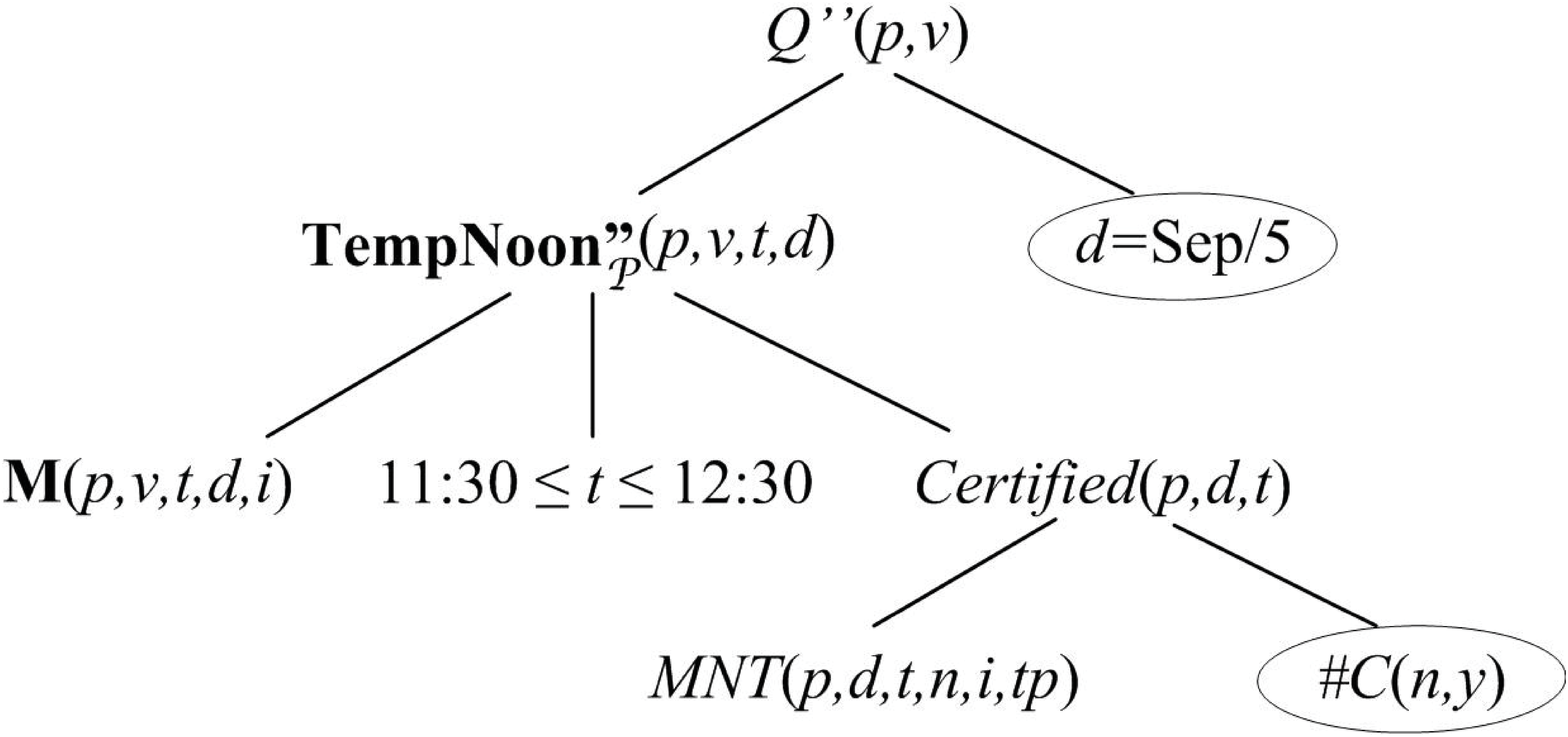,width=90mm}
    \caption{Parse tree for $\mathcal{Q}''(\nit{p, v})$}
    \label{fig:initialtree}
    \vspace*{-1mm}
\end{figure}

\begin{example}  \label{exa:queryexternal2} (example \ref{exa:queryexternal} continued) \
In order to obtain quality answers to query (\ref{eq:query0}), the
query  is  rewritten, as expected,  as
\begin{equation}
\mathcal{Q}''(\nit{p, v}) \ \leftarrow \
\textbf{TempNoon''}_{\!\mc{P}}\nit{(p, v, t, d)}, \ d= \mbox{Sep/5}.
\label{eq:query2} 
\end{equation}
If we do a naive bottom-up evaluation, we will need to invoke $\nit{\#\!C}$ for
every nurse in $\nit{MNT}$ as a bulk, which, apart from not being allowed by the source, could
generate a huge amount of data.

Instead, by looking at the evaluation tree shown in
Figure~\ref{fig:initialtree} for this Datalog query  (each branching point corresponds to a conjunction), we can see that
we would rather  ask about nurses who were
working on Sept/5.

We can also see from the query tree that the nurse name $n$ is not yet bound at the
time of triggering the request for external data. In consequence, we have to make sure that both
the date  $d= \mbox{Sep/5}$ and the name are evaluated before we access the external source.
\boxtheorem
\end{example}

Top-down techniques use the value selection in the query to
restrict the evaluation to facts that are relevant to the query.
However, top-down methods are less appropriate that bottom-up
approaches when a possibly large amount of data is involved
\cite{CGT90}. By using the magic sets technique, the original
Datalog query is rewritten into a new Datalog program, the {\em
magic program}, for which the bottom-up evaluation focuses only on
relevant facts (for the original query).

We will briefly summarize and use one of the variants of this
 methodology, the ``generalized
supplementary magic" \cite[sec. 13.2-13.3]{AHV95}.
The construction of the magic program is based on the notion of
{\em sideways information passing} (SIP), according to which values are
passed from one predicate to the next (from left to right in a rule body) during
the evaluation of a rule body.

We begin by ``adorning"
each occurrence of a predicate in a rule of the original program
with a {\em label} for each variable (or argument). Such a label
indicates whether the variable is either \emph{bound} ($b$) or
\emph{free} ($f$) at the time of the SIP evaluation. Each sequence
of labels ($f$ or $b$) for a predicate is  called an
\emph{adornment}. We describe the annotation process by means of
our running example.

\begin{example}  (example \ref{exa:queryexternal2} continued) \label{exa:queryexternalMS}
The query in (\ref{eq:query2}) is an open query,
i.e. about all the tuples $(p,v)$ that satisfy the rule
body. Then, the variables $p, v$ in the query predicate $\mathcal{Q}''$
are implicitly free ($f$); and the top predicate becomes annotated
as $\mathcal{Q}''^{f\!f}$.

The other predicates that appear in the
program computing the query are also adorned, and the adorned
versions of rules in Example~\ref{exa:queryexternal} are as
follows:
\begin{eqnarray}
\mathcal{Q}''^{f\!f}(\nit{p, v}) &~\leftarrow~& d= \mbox{Sep/5},
\textbf{TempNoon''}_{\!\!\mc{P}}^{\nit{fffb}}\nit{(p, v, t, d)}.
\label{eq:query2ad} \nonumber \\
\textbf{TempNoon''}_{\!\!\!\mc{P}}^{\!\nit{fffb}}\!(p, v, t, d)
&~\leftarrow~& \nit{\textbf{M}(p, v, t, d, i)}, \ \mbox{11:30} \leq  t \leq
\mbox{12:30},\\
&&\nit{Certified^{\!bbb}\!(p,d,t)}. \label{eq:prime2ad} \nonumber \\
\nit{Certified^{\!bbb}\!(p,d,t)} &~\leftarrow~&
\nit{MNT(p,d,t,n,i,tp)}, \nit{\#\!C^{\!bf}\!(n,y)}.
\label{eq:uno2ad} \nonumber
\end{eqnarray}
In the first rule, the selection for $d$ makes it become adorned
with $b$ in the right-most predicate. The second rule is
introduced with those adornments in the head, because that is the
predicate we have to evaluate at the end of the body in the
previous rule.

In rule (\ref{eq:prime2ad}), and in general,  the extensional
predicate is not annotated since the totality of data is just there, to be used
without restriction.\footnote{We are not dealing in this paper with the optimization of access
 to given contextual instances.} The last predicate in it has all its
arguments adorned with $b$, because they all appear in a predicate
to the left in the same rule body, whose evaluation will make them
take specific values that will be used when the right-most predicate is processed.\footnote{We are
assuming that atoms are evaluated from left to right in a
body.}

The third rule is introduced to evaluate the last
predicate in  rule (\ref{eq:prime2ad}). As before, the right-most
predicate in the body has its first argument, $n$, adorned with
$b$, because its appears in a predicate to the left.
 \boxtheorem
\end{example}

The process we just illustrated begins by adorning the query predicate (its
head), and next its body. The bindings are then propagated
to the other rules. Only intentional predicates
are adorned; and we consider external predicates as special cases of
intentional predicates since we do not have explicit extensions for them.

The adornment process is the first step. The resulting adorned program
 has now to be rewritten in terms of the so-called {\em magic
predicates}, which is done as follows. For each rule $p \leftarrow
a_1, \ldots, a_m$ in the adorned program (excluding the query), we
first create a new predicate $magic\_p$ containing only the bound
arguments of $p$. Next,  a supplementary relation
$\nit{sup}^p_i$ for each predicate $a_i$ is created. The arguments in
$\nit{sup}^p_n$ are those in the rule head. The arguments in
$\nit{sup}^p_i$, with $i \neq m$, contain all the arguments that
occur in $\nit{magic}\_p, a_1, \ldots, a_i, a_{i+1}, \ldots, a_m,
\nit{sup}^p_m$.

The rule above that defines $p$ is rewritten in terms of
$\nit{magic}\_p$ and $\nit{sup}^p_1, \ldots,$ $\nit{sup}^p_m$, in such a
way that can be used to simulate a SIP strategy with a bottom-up
evaluation.\footnote{Predicates $\nit{sup}^i_j$ are also denoted with
$\nit{supmagic}^i_j$ in the literature.}

\begin{example}  \label{ex:magic} (example \ref{exa:queryexternalMS} continued) \ In the previous example we
left having to
evaluate predicates $\textbf{TempNoon''}_{\!\mc{P}}^{\nit{fffb}}$ and
$\nit{Certified^{bbb}(p,}d,t)$ in the heads of the last two
rules, resp. For this, we rewrite the rules for them
using the magic and supplementary predicates, as follows:
\begin{eqnarray}
\hspace*{-4mm}\nit{\textbf{TempNoon''}}_{\mc{P}}^{\nit{fffb}}(p, v, t, d) &~\leftarrow~&
\nit{magic\!\_\textbf{TempNoon''}}_{\!\mc{P}}^{\nit{fffb}}(d), \ \nit{\textbf{M}(p, v, t,
d,
i)}, \nonumber \\
& & \nit{sup}^1_1(d,p,t), \ \mbox{11:30} \leq  t \leq
\mbox{12:30}, \nit{sup}^1_2(d,p,t),\nonumber \\
& &  \nit{Certified^{bbb}(p,d,t)}, \ \nit{sup}^1_3(p,v,t,d).\nonumber
\end{eqnarray}

\begin{eqnarray}
\nit{Certified^{bbb}(p,d,t)} &~\leftarrow~&
\nit{magic\!\_Certified^{bbb}(p,d,t)}, \nit{MNT(p,d,t,n,i,tp)}, \nonumber \\
& & \nit{sup}^2_1(p,d,t,n), \nit{\#\!\!C^{bf}(n,y)},  \nit{sup}^2_2(p,d,t). \nonumber
\end{eqnarray}
According to our SIP strategy, predicates are evaluated from left
to right. Consequently,
$\nit{magic\!\_\textbf{TempNoon''}}_{\!\mc{P}}^{\nit{fffb}}$ will
be initialized with the value $Sep/5$ taken from the query.

Next, considering
the predicates by pairs, $\nit{sup}^1_1$ will receive the result
of the evaluation of $\nit{magic\!\_\textbf{TempNoon''}}_{\!\mc{P}}^{\nit{fffb}}$ and
$\textbf{M}$. In its turn, $\nit{sup}^1_2$ will get the result of the
evaluation of  $\nit{sup}^1_1$ and $\mbox{11:30} \leq t \leq
\mbox{12:30}$.  Finally, $\nit{sup}^1_3$ will receive the values
from the evaluation of $\nit{sup}^1_2$ and
$\nit{Certified^{bbb}}$.

Considering all this, we can rewrite the rule defining
$\nit{\textbf{TempNoon''}}_{\!\mc{P}}^{\nit{fffb}}$ into a set of rules:
\begin{eqnarray}
\nit{magic\!\_\textbf{TempNoon''}}_{\!\mc{P}}^{\nit{fffb}}(d) &~\leftarrow~&
d=Sep/5. \nonumber \\
\nit{sup}^1_1(d,p,t) &~\leftarrow~& \nit{magic\!\_\textbf{TempNoon''}}_{\!\mc{P}}^{\nit{fffb}}(d), \nit{\textbf{M}(p, v, t, d, i)}. \nonumber \\
\nit{sup}^1_2(d,p,t) &~\leftarrow~& sup^1_1(d,p,t), \mbox{11:30} \leq  t
\leq \mbox{12:30}. \nonumber \\
\nit{sup}^1_3(p,v,t,d) &~\leftarrow~& sup^1_2(d,p,t), \nit{Certified^{bbb}(p,d,t)}. \nonumber \\
\nit{\textbf{TempNoon''}}_{\!\mc{P}}^{\nit{fffb}}(p, v, t, d) & \leftarrow &
\nit{sup}^1_3(p,v,t,d). \nonumber
\end{eqnarray}
The rewriting for $\nit{Certified^{bbb}}$ is similar:
\begin{eqnarray}
\hspace*{-6mm}\nit{magic\!\_Certified^{bbb}(p,d,t)} &~\leftarrow~& \nit{sup}^1_2(d,p,t). \nonumber \\
\nit{sup}^2_1(p,d,t,n) &~\leftarrow~& \nit{magic\_Certified^{bbb}(p,d,t)}, \nit{MNT(p,d,t,n,i,tp)}. \nonumber \\
\nit{sup}^2_2(p,d,t)   &~\leftarrow~& \nit{sup}^2_1(p,d,t,n), \nit{\#\!\!C^{bf}(n,y)}, \nonumber \\
\nit{Certified^{bbb}(p,d,t)} &~\leftarrow~& \nit{sup}^2_2(p,d,t).\nonumber
\end{eqnarray}
Some predicates are redundant. In general, a supplementary
predicate that appears with no other predicate in a rule's body
can be replaced by its definition. With that simplification, the
resulting \emph{magic program} for the original adorned rules of
Example~\ref{exa:queryexternalMS} becomes:

\begin{eqnarray}
\nit{magic\!\_\textbf{TempNoon''}}_{\!\mc{P}}^{\nit{fffb}}(d) &~\leftarrow~&
d=Sep/5. \nonumber\\
\nit{sup}^1_1(d,p,t) & ~\leftarrow~ & \nit{magic\!\_\textbf{TempNoon''}}_{\mc{P}}^{\nit{fffb}}(d), \ \nit{\textbf{M}(p, v, t, d, i)}. \nonumber\\
\nit{sup}^1_2(d,p,t) & ~\leftarrow~ & \nit{sup}^1_1(d,p,t), \mbox{11:30} \leq  t
\leq \mbox{12:30}. \nonumber \\
\textbf{TempNoon''}_{\!\mc{P}}^{\nit{fffb}}(p, v, t, d) & ~\leftarrow~ & \nit{sup}^1_2(d,p,t), \nit{Certified^{bbb}(p,d,t)}. \nonumber \\
\nit{magic\!\_Certified^{bbb}(p,d,t)} & ~\leftarrow~ & \nit{sup}^1_1(d,p,t),
\mbox{11:30} \leq  t \leq \mbox{12:30}. \nonumber
\end{eqnarray}

\begin{eqnarray}
\hspace*{-4mm}\nit{sup}^2_1(p,d,t,n) & ~\leftarrow~ & \nit{magic\!\_Certified^{bbb}(p,d,t)}, \ \nit{MNT(p,d,t,n,i,tp)}. \nonumber \\
\nit{Certified^{bbb}(p,d,t)} &~\leftarrow~& \nit{sup}^2_1(p,d,t,n), \ \nit{\#C^{bf}(n,y)}, \nonumber\\
\mathcal{Q}^M(\nit{p, v}) &~\leftarrow~& d= \mbox{Sep/5}, \
\textbf{TempNoon''}_{\!\mc{P}}^{\nit{fffb}}(p, v, t, d). \nonumber
\end{eqnarray}
The last rule collects the (quality) answers to the query in (\ref{eq:query2}).

The magic query program has the evaluation tree in
Figure~\ref{fig:magictree}. We can use and follow it for a bottom-up
evaluation of the query (predicate) at the root.

\begin{figure}
\vspace*{-1mm}
    \centering
       \epsfig{file = 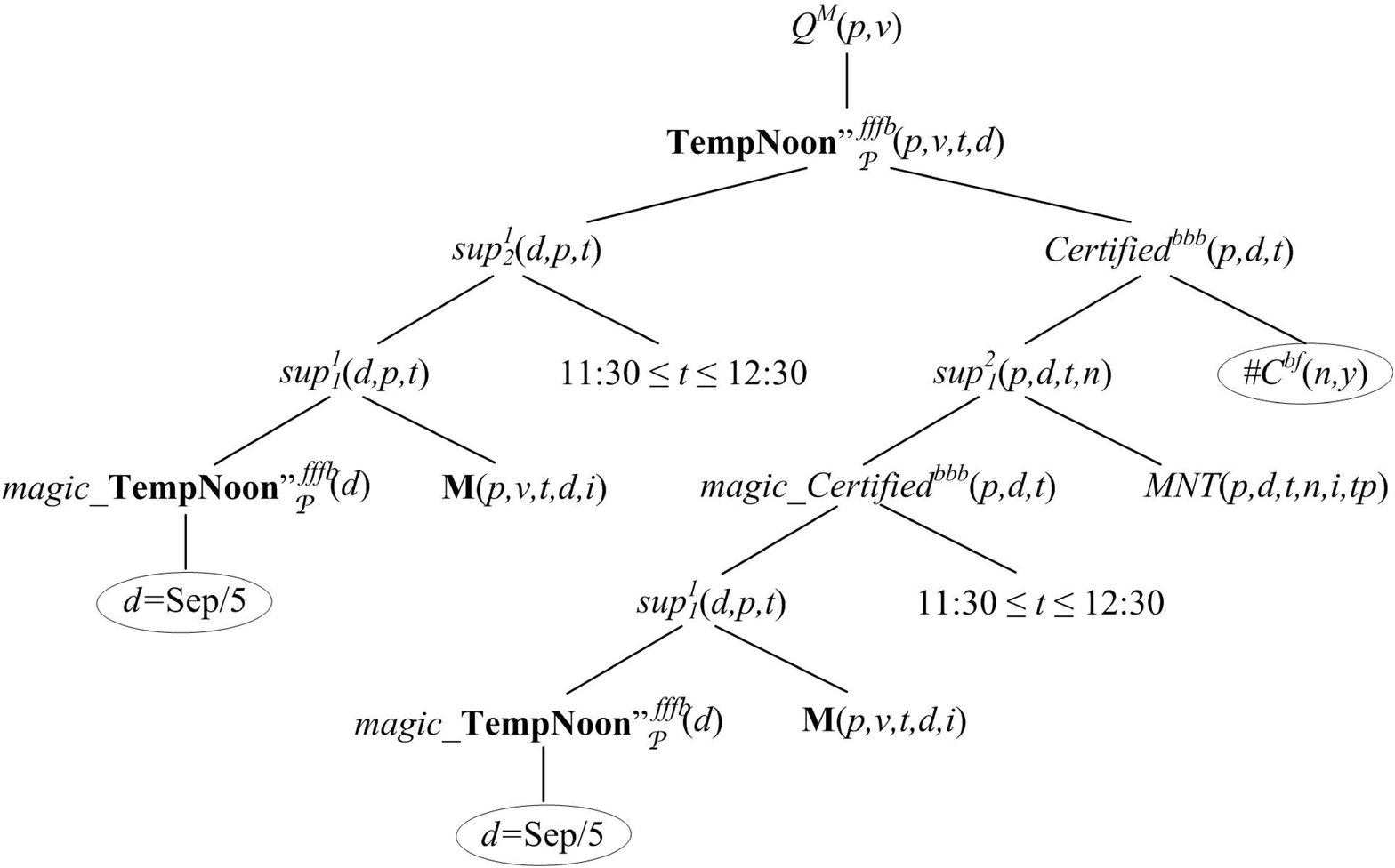,width=125mm}
    \caption{Parse tree for $\mathcal{Q}^M(\nit{p, v})$}
    \label{fig:magictree}
    \vspace*{-1mm}
\end{figure}

This
magic program and its bottom-up evaluation will take full advantage of the selections involved.
Notice that, by
the time we need to issue a data request to the external predicate $\#\!C$, the value
selection $d=Sep/5$ has already been applied. That is, the
external predicate will receive a restricted subset of names,
those of the nurses that were working on $Sep/5$.

In fact, if we consider the
subtree rooted at $\nit{Certified}$, we notice that the value selection $d=Sep/5$
propagates upwards, and therefore $sup^1_1$,
$\nit{magic\!\_Certified}$, and $sup^2_1$ all receive only tuples
corresponding to $Sep/5$. When the request for external data is
issued, only three names are left: Susan, Cathy and Joan. Since predicate $\#C$ has the adornment $\#C^{bf}$,  a specific
query
to $\#C$ for each of these 3 values is issued. In
contrast, an evaluation based on the tree in
Figure~\ref{fig:initialtree} would send a query to $\#\!C$ for all the nurse names
appearing in relation $\nit{MNT}$.
\boxtheorem
\end{example}

Having the possibility of taking into account the value selections
involved in the query when evaluating it bottom-up is crucial for
dealing with external sources. We want to be as restrictive as
possible when querying them, while still producing the correct
results.

\ignore{
\comlb{OLD: We should say here that in the definition
(\ref{eq:uno2}), it is most likely that the external predicate
$\#C$ needs the binding $\#C^{bf}$, i.e. ..., because .... In the
previous example, after applying the MS, it happened that the
predicate got the right binding. However, this may not be the case
in general. So the question is: How can we make sure that this
happens? Can we accommodate/change the MS method to do this? A
general methodology? MS guided by the initially given bindings
(too)?}  }

Notice that in definition
(\ref{eq:uno2}), we would expect to call $\#C$ always with a specific value for $n$. In other words,
predicate $\#C$ should always  have the binding $\#C^{bf}$. In the
previous example, after applying the magic set (MS) methodology, it happened that this
predicate got the right binding, obtaining a {\em safe} call to $\#C$. However, this raises the issue
as to whether we can always guarantee that this will happen (possibly through a modified MS
that is also guided by the initially given bindings).

\ignore{
\comfr{This has been addressed in the next paragraph and with the
rewriting described in Subsection~\ref{sec:triggers}}  }

We addressed this problem by assuming the
{\em input guarded condition}. It  guarantees that the
right bindings for the external predicates are maintained after
applying MS. More precisely, if the rule is input guarded, then every input
variable of the external predicate $\#E$ also appears in some atom
before it. Thus, by applying MS to a rule $p \leftarrow a_1,
\ldots, a_n, \#E, a_{n+1}, \dots, a_m$, we produce a set of rules
for $\nit{sup}^p_i$ that maintain
the bindings. The reason for that is that the arguments of the
$\nit{sup}^p_i$ contain the variables that appear before and after
them in $p$. In particular, they will contain the input variables
of $\#E$.

\ignore{
\comlb{OLD[ER]: Here we should illustrate (and then develop) the
idea of having triggers requesting the right external data. Or
maybe in a separate section?}
\comlb{OLD:About previous comment: we could illustrate the issue
right away, as a continuation example, asking for the
certification for Susan via a trigger. I would not go much beyond
that ... That is, we should have a trigger that has the first
argument as an input parameter.}
\comfr{Addressed in Example \ref{exa:triggers}.}
}



Once we have the MS rewriting, the question is how to get data
from the external sources given their access restrictions. We have
seen how the input guarded condition guarantees the right bindings
in terms of input/output parameters.

Our evaluation methodology has to implement the right
access method, one  that is compatible with the access restrictions
as expressed via input/output parameters. This can be done by placing  a {\em trigger} or {\em active rule} \cite{adbs97} in a DBMS at the contextual level, or through an application program running from the context. Either way,
a {\em procedure} is needed  that request the output values from external for given input values.

For example,  we may have a defining rule
containing an external source, say
\begin{eqnarray}
\hspace*{1cm}p(\bar{x}) &~\leftarrow~& a_1(\bar{x}_1), \ldots, a_j(\bar{x}_j),
\#E(i_1,\ldots,i_n,o_1,\ldots,o_m), \label{eq:external}\\
&& a_{j+1}(\bar{x}_{j+1}), \dots,
a_k(\bar{x}_k), \nonumber
\end{eqnarray}
where $i_1,\ldots,i_n$ are the input parameters
and $o_1,\ldots,o_m$ are the output parameters. Notice that $i_1, \ldots, i_n$ appear
in $\cup_{i=1}^j \bar{x}_i$.

In this case, we create a procedure $\nit{get\#\!E}[I_1,\ldots,  I_n;O_1, \ldots, O_m]$, with the obvious
input/output parameters. We may assume that the output values are \nit{null} when the procedure is undefined for a particular combination of input values. Furthermore, assuming a SIP evaluation, we define a new predicate
$\nit{input(i_1,\ldots,i_n)}$ by: \ $\nit{input}(i_1,\ldots,i_n) \ \leftarrow \ a_1(\bar{x}_1), \ldots,
a_j(\bar{x}_j)$.

Every time a new tuple $\nit{input}(c_1,\ldots,c_n)$ is created,
while
evaluating (\ref{eq:external}), the procedure is invoked
as $\nit{get\#\!E}[c_1,\ldots,  c_n;O_1, \ldots, O_m]$. The returned values are
passed over to the predicates to the right.

\begin{example} \label{exa:triggers} (example \ref{ex:magic} continued) \
In the magic program, the external predicate appears in
the definition of $\nit{Certified}$:
\begin{equation*}
\nit{Certified^{bbb}(p,d,t)} \ \leftarrow \ \nit{sup}^2_1(p,d,t,n), \
\nit{\#C^{bf}(n,y)} \label{ex:certified}
\end{equation*}
From this we define: \
$\nit{input}(n) \ \leftarrow \  \nit{sup}^2_1(p,d,t,n)$; 
and create the procedure $\nit{get\#\!C}[N;Y]$.

\ignore{
The last three rules will replace (\ref{ex:certified}) in the
magic program.

We define then an external access trigger for the new rule
containing the external predicate, i.e., rule
(\ref{ex:external_1})

\begin{tabbing}
\hspace{3mm} \= \hspace{3mm} \= \hspace{3mm} \= \\
\> CREATE TRIGGER $input\#C$ \\
\> AFTER INSERT ON $\nit{input(n)}$ \\
\> FOR EACH $\#C.max\_tuples()$ ROWS \\
\> \> $C(n,y) \leftarrow input(n), \#C.get(n)[y]$ \\
\end{tabbing}
Assuming that $\#C.max\_tuples()=1$, t  }

In this example, the procedure will be invoked once for each nurse
name appearing as a value for $\nit{input}$, in this case, those who worked on
Sep/5, i.e. Susan, Cathy and Joan. Assuming that the (now) external source $\#C$ contains the
information in Table~\ref{tab:cert}, the external calls will
produce $\nit{get\#\!C[Susan;1996]}$,
$\nit{get\#\!C[Cathy;2009]}$, and $\nit{get\#\!C[Joan;null]}$.
\boxtheorem
\end{example}


\ignore{
\newpage

\appendix
\section*{Appendices}
\section{Quality Relaxation}
What happens when CQPs cannot be evaluated? There might be several
reasons for that.
\begin{enumerate}
\item Some contextual data may be missing and that prevents the
evaluation with some meaningful result. For instance, the LCI
restricted to the CQP might be empty. Consider the CQP
$\nit{Certified(p,d,t)}$ (\ref{eq:uno}): if the instance in
Table~\ref{tab:cert} (corresponding to $C(n,y)$) is missing, the
LCI restricted to \emph{Certified} will always be empty. \item An
external predicate is not available at evaluation time. Consider
again $\nit{Certified(p,d,t)}$ (\ref{eq:uno}). If we replace
$C(n,y)$ by an external predicate $\#C(n,y)$, whenever this latter
is not available the evaluation of \emph{Certified} will be empty.
\item There may be more than one external predicate involved and
their source descriptions may be incompatible, i.e., the CQP has
one or more non-queryable relations, i.e., a relation that cannot
be access from any instance~\cite{CM08}. The data in such
non-queryable relations cannot be used in the evaluation of the
respective CQP. \item Even when the LCI restricted to the CQP
contains in fact some data, it might be not enough to produce a
meaningful result. For instance, if $C(n,y)$ has incomplete
information and there are many certified nurses in the hospital
that do not appear in $C(n,y)$, (or if the external source provide
certification information from just a region rather than the
entire country), many quality tuples (i.e., tuples satisfying the
CQP) will not appear in the quality instance.
\end{enumerate}

Case (4) above is hard because we do not have a way to know when
the data in a source is incomplete w.r.t. the specification. (1),
(2), and (3), in contrast, can be viewed as special cases of
\emph{query relaxation}. In (1) and (2) we have the option of
relaxing the quality by dropping the entire CQP whose LCI is empty
or just not evaluating the actual rule/s that are empty. In the
example, we could eliminate either $C(n,y)$ from the definition of
$\nit{Certified(p,d,t)}$, or $\nit{Certified(p,d,t)}$ from the
definition of $TempNoon$' in (\ref{eq:prime}). In case (3), the
relaxation would involve eliminating all external predicates that
are non-queryable from the CQP definitions.

A different situation arises when the predicates can be modified
without completely discarding them. Similarly to Gaasterland et
al. \cite{Minker92} we can think of two types of relaxation in
such cases: predicate generalization and breaking a join
dependency. (Here we consider broadening the domain of a variable,
as presented in \cite{Minker92}, as a special case of predicate
generalization.)

\subsection{Predicate generalization}

We can assume a hierarchy of predicates, from more specific ones
to more general ones. The more specific a predicate, the more data
quality it provides. For instance, $C(n,y)$ with the name and year
of the nurses' certification can be part of a hierarchy that
includes $\nit{Coll(n,y)}$ (for nurses with college degrees) and
$\nit{Univ(n,y)}$ (for nurses with university degrees). Such a
quality hierarchy would be given as follows:

\[ \nit{Coll(n,y)} \prec \nit{Univ(n,y)} \prec C(n,y) \]

where $C(n,y)$ is the most specific (and desirable) predicate and
$Coll(n,y)$ most general (and least desirable) one.

For instance, consider again $\nit{Certified(p,d,t)}$
(\ref{eq:uno}). If we replace $C(n,y)$ by $\nit{Univ(n,y)}$ we
obtained the following relaxed expression:

\[ \nit{Certified^{rel}(p,d,t)} \leftarrow \nit{Temp(p,d,t,n,i,tp)},
\nit{Univ}(n,y) \]

which now has the lower quality requirement of a university degree
rather than a certification.

A similar situation occurs when we want to broaden the domain of a
variable. For instance, the definition of $TempNoon$ in
(\ref{eq:prime}) has a range predicate on $t$ specifying an
interval [11:30,12:30]. A relaxation by broadening the valid
domain of $t$ would entail changing that interval to [11:00,1:00].
This change would still satisfies the condition ``around noon'',
although in a broader sense.

Consider that predicate $\nit{S(d, s, n)}$ had information about
the shifts not only of nurses but of doctors as well. The original
definition of $\nit{Temp(p,d,t,n,i,tp)}$ would be as follows:

\begin{eqnarray}
\nit{Temp(p,d,t,n,i,tp)} &\leftarrow& \nit{M(p, v, t, d, i)},  \nit{S(d, s, n)},  Nurse(n), \nit{T(n, d, tp)}, \nonumber \\
 & & \nit{i}= \mbox{therm}, \mbox{4:00} < \nit{t} \leq \mbox{12:00}, \nit{s}=\mbox{morning}. \phantom{ppp} \label{eq:nursesAMnew}
\end{eqnarray}

Remember that original requirement was that nurse had to take the
temperature of patients, so the predicate $Nurse(n)$ was added to
restrict the domain of $n$. We could relax that requirement by
changing the domain of $n$ from $Nurse(n)$ to $Medical(n)$, a
category containing all medical stuff, both nurses and doctors.
The resulting predicate will be

\begin{eqnarray}
\nit{Temp(p,d,t,n,i,tp)} &\leftarrow& \nit{M(p, v, t, d, i)},  \nit{S(d, s, n)},  Medical(n), \nit{T(n, d, tp)}, \nonumber \\
 & & \nit{i}= \mbox{therm}, \mbox{4:00} < \nit{t} \leq \mbox{12:00}, \nit{s}=\mbox{morning}. \phantom{ppp}
\end{eqnarray}

\subsection{Breaking a join dependency}

Another type of relaxation involves weakening or breaking a join
dependency. For instance, in expression (\ref{eq:nursesAMnew})
above, there is a join dependency between $\nit{S(d, s, n)}$ and
$\nit{T(n, d, tp)}$ which basically states that the temperature is
to be taken by the nurse in charge of the shift. However, nothing
prevents other nurses, or even a doctor, to take a good
measurement of the patient's temperature. If we want to allow such
a case, we could break the dependency by renaming the variable $n$
in the second predicate. The resulting expression would be the
following:

\begin{eqnarray}
\nit{Temp(p,d,t,n,i,tp)} &\leftarrow& \nit{M(p, v, t, d, i)},  \nit{S(d, s, n)},  Nurse(n), \nit{T(n', d, tp)}, \nonumber \\
 & & \nit{i}= \mbox{therm}, \mbox{4:00} < \nit{t} \leq \mbox{12:00}, \nit{s}=\mbox{morning}. \phantom{ppp}
\end{eqnarray}
}

\ignore{ POSTREFERENCES!!!!

}

\end{document}